**Direct Trajectory Reconstruction for Fast 3D X-ray Particle Tracking Velocimetry in Porous Media**

R. van der Merwe[1,2,a)], W. Goethals[1,2,3], S. Ellman[1,2], S. Manoorkar[1,2], J. Aelterman[1,4], M. N. Boone[1,3], and T. Bultreys[1,2]

[1)]Centre for X-ray Tomography, Ghent University, Proeftuinstraat 86, 9000 Ghent, Belgium
[2)]Department of Geology, Ghent University, Krijgslaan 281, 9000 Ghent, Belgium
[3)]Department of Physics and Astronomy, Ghent University, Proeftuinstraat 86, 9000 Ghent, Belgium
[4)]Department of Telecommunications and Information Processing – imec, Ghent University, Sint-Pietersnieuwstraat 41, 9000 Ghent, Belgium

## Abstract

X-ray tomographic particle tracking velocimetry is a promising tool to provide pore-scale insight into fluid flow inside opaque porous media. This is critical for applications ranging from geo-energy to electrochemistry. Current approaches to implement this three-dimensional, three-component (3D-3C) velocimetry technique use conventional micro-computed tomography (CT) workflows to recover tracer particle trajectories. These workflows first reconstruct a sequence of 3D CT images from radiographs which are sequentially acquired at time intervals that span multiple viewing angles. The tracer particles are then detected in each of these CT images and finally linked into trajectories. However, motion-blurring induced by the static CT volume reconstruction impairs the detectability of fast-moving tracer particles, severely limiting the temporal resolution and dynamic range of the velocimetry outcome. In this work, we show that this limitation can be overcome by explicitly recovering particle trajectories directly from the acquired radiographs, bypassing the intermediate tomographic reconstruction step and the static-scene assumptions of traditional 3D CT reconstructions. We implement this approach in a new open-source algorithm, CTracks, and validate it using experimental data on pipe flow and on flow through a complex porous medium, demonstrating a five-fold increase in the maximum detectable velocity compared with state-of-the-art methods. Numerical simulations further confirm improved detection rates and bounded trajectory errors at these increased velocities. Our work thus demonstrates that 3D-3C velocimetry in opaque microscopic geometries can be extended to substantially faster flows by replacing X-ray tomographic image reconstruction with direct particle trajectory reconstruction.

---

[a)]Corresponding author, Robert.vanderMerwe@UGent.be

# I. INTRODUCTION

Fluid flow in porous media plays a central role in natural and engineered systems, including applications in geological carbon sequestration (Bachu, 2015; Bui et al., 2018), subsurface hydrogen storage (Muhammed et al., 2022), groundwater remediation (Bear & Cheng, 2010), and polymer electrolyte fuel cells (Mularczyk et al., 2020, 2021). Their intricate pore geometries lead to complex, multiscale flow phenomena that remain difficult to explain and predict. Macroscopic continuum models rely on constitutive relationships which must appropriately reflect the pore-scale fluid dynamics (Blunt, 2017). Developing and validating such relationships remains challenging for many flow phenomena, including multiphase (Armstrong et al., 2016; Berg et al., 2026), non-Newtonian (Browne & Datta, 2021; Datta et al., 2022) and turbulent flows, as the governing pore-scale dynamics are not yet sufficiently understood. There is therefore a need for *in situ* experimental techniques capable of directly resolving three-dimensional pore-scale flow fields while maintaining micrometre-scale spatial resolution and sensitivity to transient flow dynamics.

While optical velocimetry techniques are highly effective in transparent micromodels and refractive-index-matched systems (Lu et al., 2018; Roman et al., 2016; Schanz et al., 2016), their reliance on optical access limits their applicability in real porous materials, whose chemical composition, pore geometry and microscopic surface roughness can fundamentally alter flow behaviour. For real, opaque porous media, X-ray micro-computed tomography (μCT) has become a key experimental tool for non-destructive three-dimensional imaging of pore structure and fluid distributions (Berg et al., 2013; Wildenschild & Sheppard, 2013). Extending these imaging capabilities from structural characterisation to direct measurements of flow kinematics has motivated the development of tomographic X-ray particle tracking velocimetry (XPTV) for non-destructive, *in situ* measurements of three-dimensional flow fields in optically opaque (pore) geometries (Bultreys et al., 2022; Mäkiharju et al., 2022). These methods build on earlier X-ray flow measurement approaches developed for simpler geometries with more coherent flow fields (Dubsky et al., 2012; Fouras et al., 2007), extending such measurements to the complex pore structures encountered in porous media.

Unlike optical particle-tracking techniques, which infer three-dimensional particle positions from a small number of simultaneously acquired views, XPTV for porous geometries typically relies on their 3D position, retrieved from a tomographic reconstruction of hundreds of radiographs acquired sequentially during sample rotation. This sequential acquisition is a consequence of the high spatial resolution required to resolve microparticles within complex porous geometries (Cnudde & Boone, 2013; Wildenschild & Sheppard, 2013), which is typically achieved using commercial micro-CT systems based on a single source-detector pair. Previous studies have demonstrated the feasibility of recovering fully three-dimensional pore-scale flow fields using a reconstruct-detect-link (RDL) workflow, in which tracer particles are detected in reconstructed CT volumes and linked into trajectories. This approach has enabled measurements in porous media ranging from creeping flows in laboratory systems (Bultreys et al., 2022), through observations of pore-clogging dynamics (Mäkiharju et al., 2022), to faster transient dynamics at synchrotron facilities (Bultreys et al., 2024).

Despite these advances, tracking microparticles at higher velocities remains a fundamental challenge. Conventional CT reconstruction assumes that the imaged system, comprising the porous matrix, fluids and tracer particles, remains effectively stationary during the acquisition of the radiographs used to reconstruct a single CT volume, hereafter referred to as a frame. If a particle moves appreciably relative to the voxel size during this acquisition period, its attenuation signal is spread over an extended region of the reconstructed volume, producing severe motion-blur artefacts. This reduces the attenuation contrast relative to the reconstruction noise floor and prevents reliable particle detection, as illustrated in the simple capillary geometry shown in Fig. 1. As a result, the maximum trackable velocity is constrained by voxel size and scan time, permitting in practice particle displacements of only about four voxels per tomographic acquisition before motion blur prevents reliable particle detection and tracking (Bultreys et al., 2022, 2024). For typical state-of-the-art laboratory XPTV acquisitions, this corresponds to velocities of less than 2 μm/s, preventing the measurement of many faster pore-scale flows.

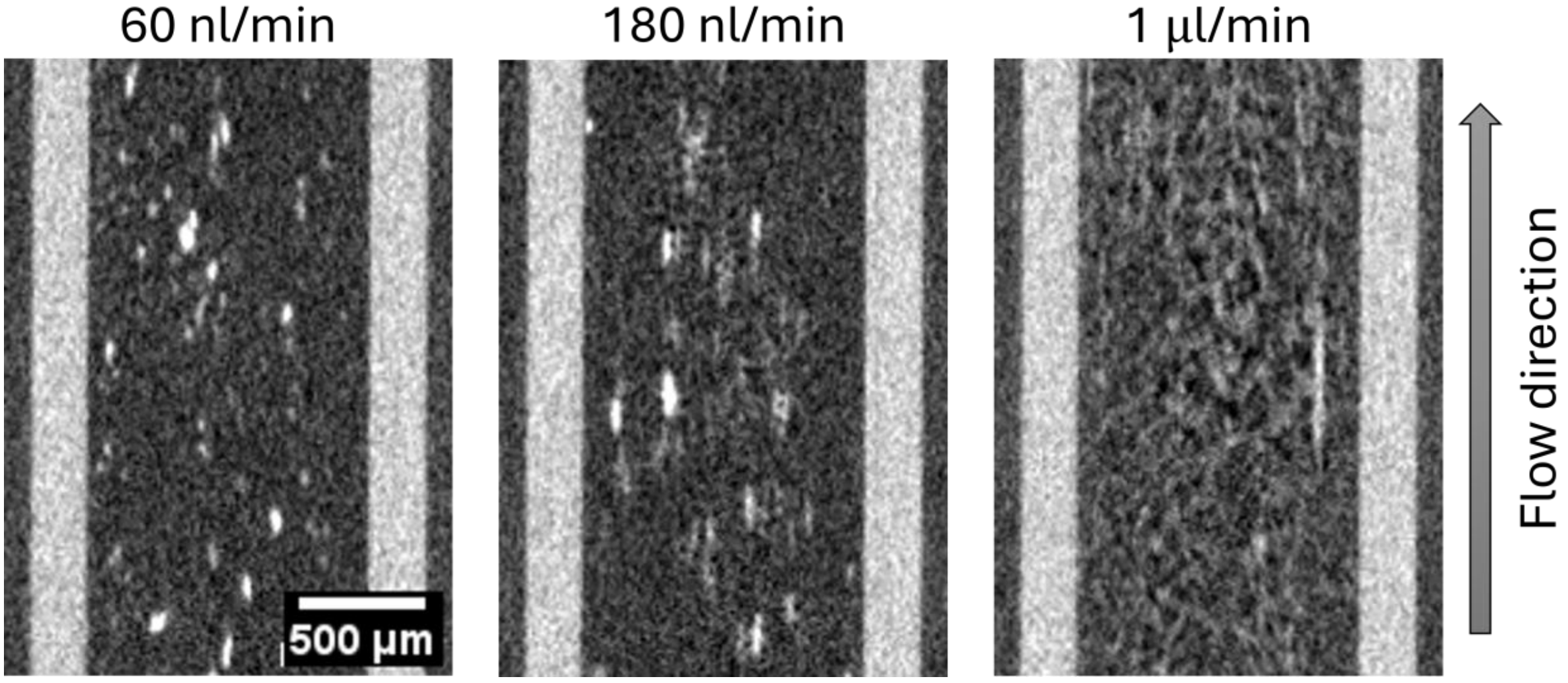

FIG. 1: Reconstructed central vertical slices of flow through a capillary at flow rates of 60, 180 and 1000 nl/min (left to right), corresponding to mean interstitial velocities of 1.1, 3.2 and 17.5 μm/s and maximum particle displacements of 4, 14 and 70 voxels per tomographic acquisition (frame), respectively. Bright spots indicate tracer particle positions. As the flow velocity increases, particle signals become increasingly motion blurred, reducing detectability and hindering tracking.

Although distinct from velocimetry, related advances have emerged in the field of dynamic X-ray tomography, which addresses the challenge of recovering time-varying information from sequentially acquired CT scans (3D “frames”). Several approaches have been proposed to overcome the temporal limitations of frame-based CT, including stroboscopic 4D X-ray microscopy (Tekseth et al., 2024) and continuous or neural dynamic reconstruction methods such as DYRECT (Goethals et al., 2025) and NeCT (Friis et al., 2025). However, these approaches do not directly address the recovery of tracer-particle trajectories required for XPTV. Stroboscopic methods require repeatable dynamics and are therefore incompatible with the non-repeatable motion of individual tracer particles. DYRECT and NeCT exploit sparsity in the temporal evolution of a dynamic attenuation field on a fixed spatial grid, making them well suited for dynamic imaging of evolving flow structures but less suited to XPTV applications that require localisation and trajectory reconstruction of small tracer particles.

To address the stringent maximum velocity limitation in XPTV, we introduce CTracks, an open-source iterative 4D tomographic reconstruction framework designed to recover particle motion beyond the detectability limits of conventional volumetric tracking. Although moving particles may no longer be detectable in reconstructed CT volumes, their attenuation signals remain encoded in the sequential radiographs acquired during a tomographic scan. CTracks therefore bypasses the intermediate CT volume reconstruction and recovers particle trajectories directly from these measurements. By incorporating a parametric model of particle motion into the tomographic forward model, particle tracking is formulated as an optimisation problem in which particle trajectories and selected acquisition parameters are iteratively refined to match the measured radiographs. While the current framework assumes a stationary pore geometry, it is not inherently limited to this assumption and could be extended to accommodate more general dynamic geometries.

In this paper, we present the development and validation of the CTracks framework. Section II describes the formulation and implementation of the iterative tracking algorithm, while Section III outlines the experimental and simulation workflows used to acquire validation data for methodology evaluation. Finally, Section IV presents the analysis of both the experimental and simulated datasets. We show that CTracks extends the attainable tracking range to velocities up to five times higher than the conventional RDL workflow, thereby recovering fast flow paths that are otherwise obscured by motion blur.

## II. METHODOLOGY

During XPTV acquisition, sequential X-ray transmission radiographs are recorded as tracer particles flow through a rotating sample. Direct trajectory reconstruction formulates particle tracking from these measurements as an inverse problem, using a particle-based forward model to recover the three-dimensional motion that best explains the measured data. Within the CTracks framework implemented here, this reconstruction takes as input the time-resolved radiograph sequence, the associated CT acquisition geometry, and a pore-space segmentation defining the accessible fluid regions. The radiographs are divided into ranges of angles over which particle motion is recovered, termed reconstruction intervals. In the present implementation each interval spans one complete CT rotation and particle motion is represented by a linear trajectory. These implementation choices are motivated in the following subsections and are not inherent requirements of direct trajectory reconstruction.

As summarised in Fig. 2, the framework comprises six stages:

A. Subtraction of the static sample background from the measured radiographs.
B. Forward projection of candidate particle trajectories to generate synthetic radiographs.
C. Iterative optimisation of the particle and acquisition parameters against measured data.
D. Management of the candidate particle population during and following optimisation.
E. Post-processing to remove spurious trajectories.
F. Interpolation of the retained velocities into a continuous pore-scale velocity field.

These stages are described in detail in the following subsections.

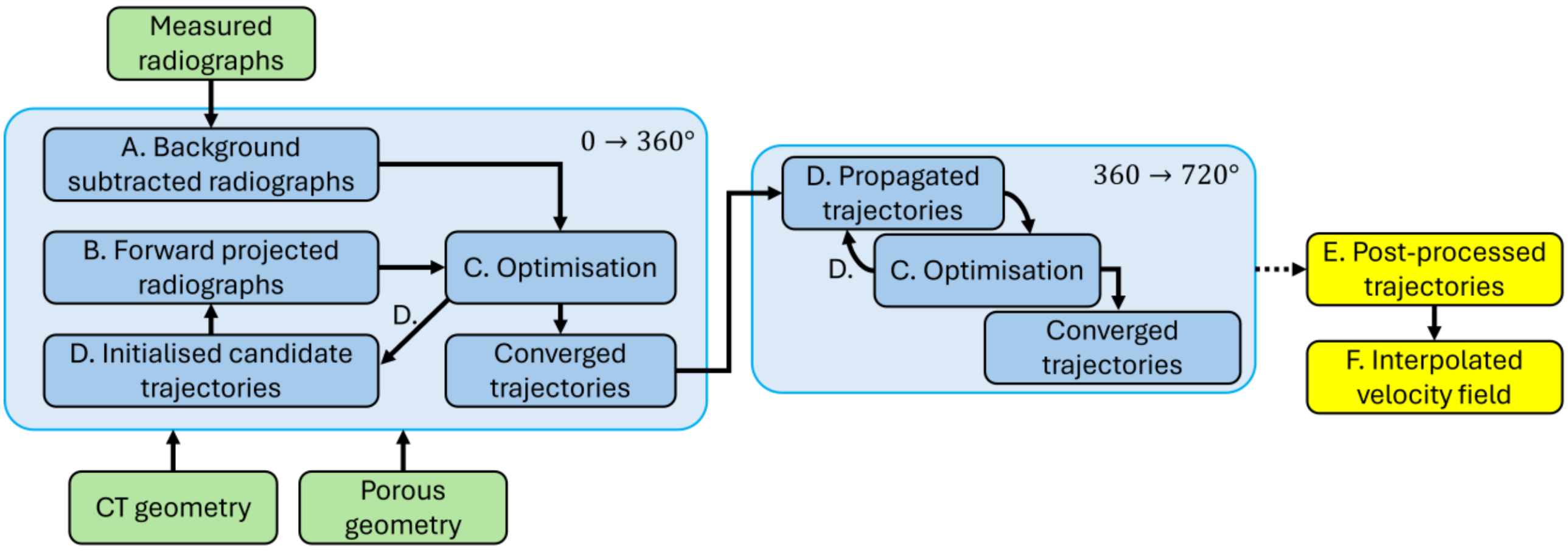

FIG. 2: Schematic overview of the CTracks workflow for direct reconstruction of particle trajectories from sequential X-ray radiographs. Green boxes indicate inputs, blue boxes represent processing steps within a reconstruction interval, and yellow boxes denote outputs. The first reconstruction interval includes particle initialisation, while subsequent intervals are initialised from particle states propagated from the previous interval. Only the first two reconstruction intervals are shown explicitly, and the workflow is repeated for all remaining intervals. Numbered stages correspond to the methodology subsections.

### A. Static background estimation and subtraction

Because the measured radiographs contain both the static sample background and the dynamic tracer particles, the first preprocessing step is the separation of the tracer-particle signal from the static sample contribution. Here, the tracer-particle signal refers to the combined attenuation contribution of all moving tracer particles present in the radiographs. The objective of this step is to remove the static sample contribution so that subsequent processing can focus on the dynamic tracer signal. In porous media, this separation is non-trivial due to the complex porous sample geometry and limited particle contrast.

Prior to background estimation, the measured radiographs are converted from X-ray transmission to optical depth using the Beer–Lambert relationship, $\tau = -\ln(I/I_0)$, where $I$ and $I_0$ represent the transmitted and incident X-ray intensities, respectively. This transformation converts multiplicative attenuation into an additive quantity, simplifying the separation of the static background and tracer-particle signals.

A particle-free background is estimated from the static attenuation contribution observed over repeated CT rotations at identical projection angles. After conversion to optical depth, all projections acquired at the same projection angle are grouped together and the minimum optical-depth value observed over time is taken as the background estimate. Because the tracer particles considered here are highly attenuating, this minimum is assumed to correspond to the attenuation of the particle-free sample. Subtracting the estimated background from each projection separates the moving tracer-particle signal from the static sample contribution. Because this approach relies on temporal background subtraction, slowly moving particles may be partially incorporated into the estimated background, reducing the particle signal available for reconstruction. In addition, estimating the background using the temporal minimum introduces a positive offset in the residual noise, as the resulting background-subtracted noise is no longer zero mean. This offset is accounted for during optimisation, as described in the Supplementary Information.

The resulting projections are shown in Fig. 3. Although background subtraction improves tracer visibility, individual particles remain difficult to identify in a single radiograph. Importantly, CTracks does not rely on the detection or segmentation of individual particles within individual projections. Instead, as radiographs are acquired sequentially during sample rotation, each particle contributes a consistent attenuation signal across multiple radiographs. These measurements can be visualised as a sinogram, a compact representation of detector intensity across projection angles. While the visualisation in Fig. 3 is primarily illustrative, it highlights how tracer-particle signals persist across successive radiographs, allowing weak particle signals to be distinguished from uncorrelated image noise and separated from the largely random measurement noise present in individual projections.

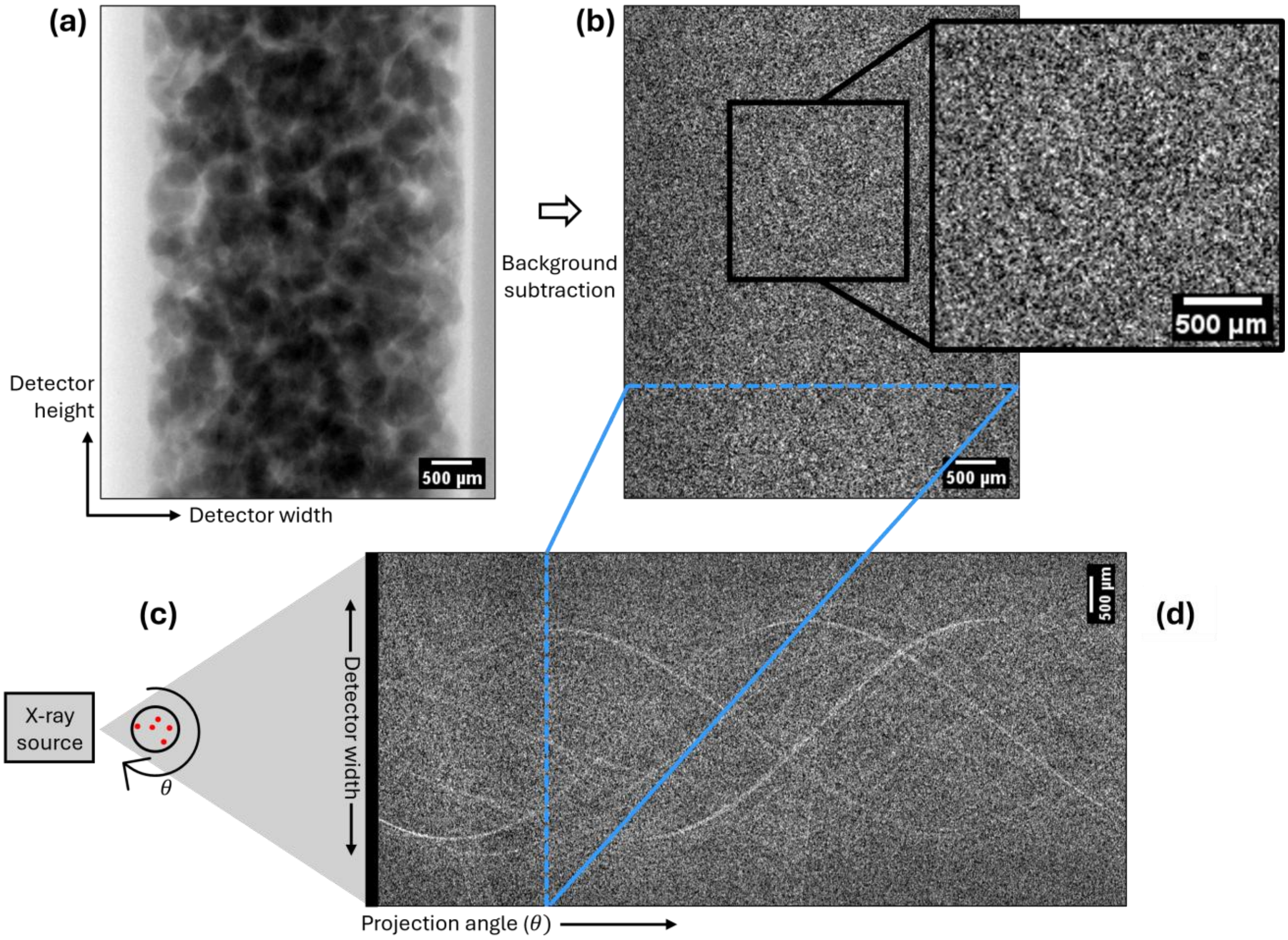

FIG. 3: Isolation of the tracer particle signal from X-ray radiographs. Raw radiographs (a) are background subtracted (b) to remove the static attenuation of the porous medium, enhancing the signal of the moving tracer particles. The inset in (b) shows a magnified view of the residual particle signal. During CT acquisition (c), radiographs are recorded sequentially as the sample rotates through projection angle $\theta$. As illustrated between (b) and (d), detector rows from successive radiographs are stacked to form a sinogram slice. While individual particles remain difficult to identify in a single radiograph, when considered over time they trace continuous, approximately sinusoidal trajectories in sinogram space (d), illustrating the temporal information exploited during direct trajectory reconstruction.

## B. Trajectory forward projection

To evaluate how well a candidate particle configuration explains the measurement, CTracks generates synthetic radiographs and compares them to the measured data within an iterative reconstruction scheme. This requires calculating the projected positions of all candidate particles at each projection time so that their attenuation signatures can be mapped onto the detector.

Each particle is modelled as a solid sphere with three-dimensional position vector $\mathbf{x} = (x, y, z)$, radius $r$ and attenuation coefficient $\mu$. Particle motion throughout a reconstruction interval is represented by a linear trajectory model $\mathbf{x}(t)$,

$$\mathbf{x}(t) = \mathbf{x_0} + \frac{t - t_0}{t_1 - t_0}(\mathbf{x_1} - \mathbf{x_0}), \tag{1}$$

where $\mathbf{x}_0$ and $\mathbf{x}_1$ denote the particle positions at the start and end of the trajectory segment, while $t_0$ and $t_1$ define the start and end times of the interval. This parameterisation assumes constant particle velocity over the interval and provides a continuous estimate of particle position at arbitrary projection times. The linear model was chosen as the simplest extension of the stationary-particle assumption used in conventional CT reconstruction, although the framework is not inherently restricted to linear trajectories.

The projected centre of particle $n$ $(u_n(t), v_n(t))$ is determined from the sample-space coordinates through the CT geometry transformation $\mathbf{g}_{\mathrm{CT}}$,

$$\big(u_n(t), v_n(t)\big) = \mathcal{V}(\mathbf{x}_n(t), \mathbf{g}_{\mathrm{CT}}, t), \tag{2}$$

where $\mathbf{g}_{\mathrm{CT}}$ contains the relevant acquisition parameters, including the centre of rotation and the source-object and source-detector distances, while $t$ determines the corresponding projection angle.

For a given projection, particle $n$ is represented on the detector as a small two-dimensional attenuation patch centred at $(u_n, v_n)$. Its optical depth is evaluated over the projected particle cross-section and is given by

$$\tau_n(u', v') = 2\mu\sqrt{(mr_n)^2 - (u')^2 - (v')^2}\,, \tag{3}$$

where $\tau_n$ is zero outside the particle boundary. The local detector coordinates $u'$ and $v'$ are measured relative to the projected particle centre and $r$ is scaled by the local cone-beam magnification factor $m$. Because the particle diameter is small relative to the source-detector distance, a small-angle approximation is used, and the magnification is assumed constant across the projected image of an individual particle.

To evaluate the particle contribution on the detector grid $(u_d, v_d)$, the coordinates are mapped relative to the projected centre of each particle. The generated projection is then formed by summing the contributions of all $N$ particles on the detector grid

$$\boldsymbol{p}_{gen}(u_d, v_d, t) = \sum_{n=1}^{N} \tau_n(u_d - u_n(t), v_d - v_n(t)). \tag{4}$$

Repeating this procedure for all projection times produces the complete generated sinograms used during optimisation.

### C. Optimisation procedure

The goal of the optimisation procedure is to identify the particle trajectories and imaging parameters that best reproduce the measured radiographs. The optimised parameters consist of the particle start and end coordinates $\mathbf{x}_{0,1}$, particle radius $r$, particle attenuation coefficient $\mu$, and the CT centre-of-rotation. Trajectory reconstruction is performed independently for each reconstruction interval. As described above, each reconstruction interval corresponds to a single CT rotation. This choice balances the amount of orthogonal information available for reconstruction against the validity of the linear-motion assumption.

For each reconstruction interval, the generated sinogram is compared with the measured sinogram using a mean-squared-error (MSE) loss over all detector pixels and projection times,

$$\mathcal{L} = \frac{1}{N_{pixels}} \sum_{t}^{N_t} \sum_{j}^{N_j} \left(p_{j,t;gen} - p_{j,t;scan}\right)^2 \tag{5}$$

Inspired by recent developments in differentiable forward projectors for tomographic reconstruction (Gopalakrishnan & Golland, 2022; Jiang et al., 2025; Koo et al., 2021), CTracks uses GPU-accelerated automatic differentiation to efficiently optimise particle and geometry parameters. During forward projection, all operations linking the optimised parameters to the generated sinogram are recorded to form a differentiable computation graph, enabling gradients of the loss with respect to $\mathbf{x_0}, \mathbf{x_1}, r, \mu$ and the centre of rotation to be computed and used for iterative gradient-based updates.

To improve computational efficiency, optimisation is performed using small batches of projections sampled from throughout the reconstruction interval rather than using all projections simultaneously. Projection angles are randomly shuffled before batching to maximise angular diversity within each optimisation step. Convergence is monitored through the evolution of the loss function, with optimisation parameters adapted as required. The specific optimisation settings used for validation are provided in the Supplementary Information.

### D. Particle population management

As described above, the particle population is treated as an additional unknown and estimated jointly with the particle trajectories. CTracks therefore combines trajectory optimisation with three particle-population management mechanisms: candidate particle initialisation, temporal propagation of reconstructed particles, and dynamic addition of particles when portions of the measured tracer signal remain unexplained.

An initial particle population is required to begin optimisation within each reconstruction interval. Candidate particle positions are initialised randomly within the accessible pore space, determined from a binary segmentation of the porous sample geometry. The initial particle population serves only as a starting estimate, with the particle count adjusted during reconstruction through the particle-population management procedures described below. Particle radius and attenuation are initialised within empirical bounds, using both prior knowledge of the tracer particles and direct investigation of the experimental projection data. Following optimisation, reconstructed particle states are propagated forward in time using the reconstructed velocity vectors. Particles located near grain boundaries are removed prior to propagation to minimise spurious trajectories. Additionally, the lowest-attenuation particles are discarded, as motivated in the following section. The propagated particle states are used to initialise the next reconstruction interval, providing an informed estimate of particle locations and promoting temporal consistency between successive trajectory segments.

To reduce sensitivity to local minima, pore-space constraints are relaxed during optimisation. Candidate particles may temporarily pass through the segmented solid phase, allowing particles that have incorrectly converged within small pores to escape and explore neighbouring pore regions where the measured attenuation data are better explained. In addition, stochastic perturbation steps inspired by Shake-The-Box (Schanz et al., 2016) are applied periodically to help candidate particles escape local minima. Following optimisation of a reconstruction interval, the trajectory endpoints are projected back onto pore-space voxels before propagation to the next reconstruction interval. This procedure assumes that the segmented pore space provides an accurate representation of the accessible flow domain.

Because the true number of tracer particles within the field of view is generally unknown, additional candidate particles are introduced when the total reconstructed X-ray attenuation, including the generated noise, is lower than the measured attenuation, indicating that part of the tracer signal is not explained by the current particle population. These particles are initialised randomly within the pore space, with their initial velocities obtained from the neighbouring reconstructed particles to promote consistency with the local flow field. The newly introduced particles are subsequently refined through the same optimisation procedure as existing particles. Together, particle initialisation, propagation, and dynamic particle insertion allow the particle population to evolve throughout the reconstruction, reducing sensitivity to the initial particle estimate while improving reconstruction completeness.

### E. Post-processing

Following optimisation, the reconstructed particle set contains both genuine and spurious tracer trajectories arising from image noise and imperfect initialisation. In practice, spurious particles are typically observed to minimise their contribution to the image-based loss by converging towards low attenuation and radius values, or by acquiring unrealistically large velocity magnitudes. While this behaviour was not explicitly imposed during optimisation, it provides a convenient mechanism for identifying and removing a significant fraction of false trajectories during post-processing.

Following optimisation over all reconstruction intervals, the recovered particle set is filtered using three stages:

- **Particle parameter filtering**: particles with exceptionally low attenuation or small radii are discarded using percentile-based thresholds. Particles with velocities near imposed optimisation bounds are also removed.
- **Local consistency filtering**: the remaining vectors are filtered for kinematic consistency using a universal outlier detection (UOD) criterion adapted from PIV/PTV analysis (Westerweel & Scarano, 2005), in which each particle is compared with a neighbourhood of nearby particles through a median-based deviation metric.
- **Isolated track removal**: particles with insufficient neighbouring trajectories within a specified search radius are removed, as their motion cannot be validated through local kinematic consistency.

The same post-processing workflow is applied to all datasets, with only the threshold values adjusted to account for differences in image noise, particle concentration and flow regime. The complete dataset-specific parameter sets used in this study are provided in the Supplementary Information.

### F. Velocity field reconstruction

Following post-processing, both CTracks and the conventional RDL workflow produce a sparse set of velocity vectors distributed throughout the pore space. While these discrete measurements enable trajectory-based analyses, comparison between tracking methods, numerical simulations and analytical expectations requires a continuous field representation.

The particle velocities are interpolated onto a continuous velocity field using natural-neighbour interpolation (Sibson, 1981). To impose an approximate no-slip condition, stationary particles are added along the grain boundaries prior to interpolation. While the solid grain volumes are not explicitly excluded during the interpolation step itself, all internal grain voxels are subsequently masked and assigned zero velocity. The resulting field is then processed using a Tikhonov-style variational regularisation (Gesemann et al., 2016). This technique minimises a cost function that penalises velocity divergence, reducing local interpolation noise while maintaining kinematic consistency for incompressible flows. This procedure produces a smooth, physically consistent pore-scale velocity field that respects both the sparse particle measurements and the complex boundary geometry.

## III. VALIDATION DATASETS

Validation of the CTracks framework was performed using both experimentally acquired and numerically simulated datasets to assess performance against analytical expectations, realistic experimental conditions and known ground truth. The experimental datasets consisted of a capillary tube and a porous glass sample. The capillary is a simple geometry where the dataset enables comparison against the analytical Hagen–Poiseuille velocity profile, while the porous glass dataset assesses performance in a realistic porous medium flow environment with heterogeneous velocities and tortuous flow paths. Nonetheless, because true particle trajectories are unknown in experiments, direct quantification of tracking accuracy is not possible in these datasets. A digital twin of the porous-glass experiment was therefore used to generate a third dataset with ground-truth particle trajectories, enabling quantitative evaluation of detection performance and trajectory reconstruction errors. Together, these datasets allow the proposed methodology to be assessed against analytical expectations, realistic experimental conditions, and known ground truth.

### A. Experimental XPTV datasets

Experimental validation was performed using XPTV measurements acquired during steady tracer-seeded flow. In an XPTV experiment, tracer particles are transported by the fluid while X-ray radiographs are recorded during continuous sample rotation. The experimental methodology broadly follows previous XPTV studies by Bultreys et al. (2022), although the key aspects relevant to the present validation are summarised here and in the Supplementary Information for completeness.

Experimental measurements were acquired for a capillary tube with an internal diameter of approximately 1.1 mm and a cylindrical porous-glass filter (ROBU P0, Germany) with a diameter of 4 mm, a height of 10 mm and mean pore size of 122 µm. All measurements were performed using a TESCAN CoreTOM micro-CT system at the Ghent University Centre for X-ray Tomography (UGCT). A compact fluidic setup was mounted directly on the rotation stage, as described by Bultreys *et al.* (2024). The working fluid was a viscous glycerol-water solution seeded with 5-22 µm diameter silver-coated hollow glass microspheres (Cospheric, USA).

For each sample, a high-quality static scan was first acquired to allow segmentation of the solid matrix, which defines the pore-space mask that is used by both tracking workflows. Additionally, a lower-quality static scan using the same acquisition settings as the dynamic measurement was acquired to facilitate background subtraction in the RDL workflow. Dynamic measurements were then performed using acquisition settings typical of time-resolved XPTV, with a voxel size of 12.2 µm and 850 projections acquired over a 30 s, 360° rotation. The acquired projections were additionally reconstructed into time-resolved CT volumes using standard filtered back projections to enable comparison with the RDL workflow.

To allow comparison between CTracks and conventional RDL methods across a range of particle displacements, measurements were acquired at volumetric flow rates ranging from 60 nl/min to 1 µl/min. These conditions span particle displacements ranging from within the conventional tracking regime to velocities expected to challenge frame-based particle detection. Depending on the flow condition, between 20 and 90 rotations were acquired to provide sufficient trajectory sampling while limiting particle settling and clogging effects during extended experiments.

### B. Simulated dataset

A digital twin of the porous-glass experiment was constructed to enable quantitative benchmarking of the tracking algorithms. The experimentally segmented porous geometry, CT acquisition geometry, and tracer-particle properties used by the reconstruction framework were also used to generate the synthetic projection data, to maximize consistency between the simulated and experimental workflows.

The segmentation of the porous glass sample obtained experimentally above was converted into a computational mesh. Steady single-phase Newtonian flow through the pore network was then simulated using OpenFOAM (Greenshields, 2026). The applied pressure gradient was chosen to produce a mean interstitial velocity approximately 30 times larger than in previous XPTV measurements, deliberately probing the performance limits of both the RDL and CTracks workflows. Ideal, massless Lagrangian tracer particles were then advected through the simulated velocity field to generate ground-truth trajectories. Particles were initially seeded throughout the pore space, with additional particles continuously injected at the inlet as the particle population was advected through the simulated flow field. Particle positions were sampled at regular time intervals matching the temporal resolution of the experimental projection acquisition.

The simulated particle trajectories were then converted into synthetic X-ray projection datasets by forward projection using the experimental acquisition geometry. Projections were generated over 15 rotations using the same acquisition geometry as in the experimental dataset. Particle size and attenuation were fixed, and Gaussian noise was added on the optical-depth values to approximate experimental imaging conditions. To avoid excessive trajectory overlap and simplify interpretation of the tracking statistics, analysis was restricted to a subset of 1000 trajectories that remained within the field of view throughout the acquisition. The resulting projections were then processed using both the RDL and CTracks workflows, enabling direct comparison between reconstructed and ground-truth trajectories.

Because ground-truth particle trajectories were available in the simulated dataset, quantitative trajectory-matching statistics could be calculated. These statistics were used throughout the simulation results to assess detection performance and reconstruction accuracy. Both the RDL and CTracks workflows represent each trajectory by a single position and velocity vector for each reconstruction interval. Accordingly, the ground-truth trajectories were reduced to an equivalent linear representation defined by their mean position and mean velocity over the corresponding reconstruction interval. A reconstructed trajectory is classified as a true positive when the reconstructed start and end points both lie within one voxel diagonal (1.73 voxels) of the corresponding ground-truth start and end points, respectively. This criterion was used for all performance metrics reported in Section IV B.

We note that the same forward projection model was used both to generate the synthetic data and to solve the inverse tracking problem. This introduces the possibility of an inverse crime (Kaipio & Somersalo, 2007), such that tracking performance on the simulated dataset may overestimate that achievable on experimental data. For this reason, the simulated validation is interpreted alongside results from experimentally acquired datasets described above.

# IV. RESULTS

## A. Experimental results

Experimental results are presented for the capillary tube and porous glass datasets introduced in Section III. The capillary experiment provides an analytical benchmark, whereas the porous-glass sample assesses performance under realistic pore-scale flow conditions. Typical reconstruction times were 5-8 min per reconstruction interval on a single NVIDIA RTX A5000 GPU, with detailed optimisation settings provided in the Supplementary Information.

### *1. Capillary tube*

We first evaluate CTracks in the capillary tube, where the expected velocity profile provides a well-defined analytical benchmark. The velocity distribution is expected to follow the Hagen-Poiseuille equation (White & Klein, 2011),

$$u(r) = \frac{\Delta P}{4\mu L}(R^2 - r^2) \tag{6}$$

where $\Delta P$, $\mu$, $L$ and $R$ represent the pressure drop, fluid viscosity, capillary length and capillary inner radius, respectively. The profile coefficient $\frac{\Delta P}{4\mu L}$ is obtained by fitting this expression to the measured radial velocity profiles.

Figure 4 shows that both algorithms recover comparable radial velocity profiles at flow rates of 60 nl/min and 180 nl/min. While the measured velocities systematically underestimate the theoretical Hagen-Poiseuille profile, the discrepancy is already apparent at the lowest flow rate, where motion-blur effects are expected to be negligible (Fig. 1), and is observed for both tracking methods. This suggests that the bias is not related to the tracking algorithms, but more likely reflects experimental uncertainties or simplifying assumptions in the analysis. However, because the bias is observed for all flow rates and both tracking algorithms, it does not affect the relative comparison between RDL and CTracks.

At the highest flow rate of 1 µl/min, corresponding to particle displacements of approximately 70 voxels per frame, the conventional RDL workflow completely breaks down. Motion blur severely degrades particle detectability, preventing reliable reconstruction of the velocity profile. In contrast, CTracks continues to recover the characteristic parabolic profile across the capillary cross-section, demonstrating a substantial extension of the measurable velocity range.

These observations are quantified in Table I. While both methods have comparable performance at the lower flow rates, the error in the fitted Poiseuille coefficient increases to 93% for RDL at 1 µl/min, whereas CTracks remains within 10% of the expected value. Similarly, CTracks maintains a close match with the expected interstitial velocity across all investigated flow rates, while RDL performs well at low and intermediate particle displacements but breaks down at the highest investigated flow rate. Together, these results demonstrate that direct trajectory reconstruction preserves both the velocity magnitude and the parabolic flow profile in regimes where conventional volumetric particle tracking fails.

Despite the previously reported practical limit of approximately 4 voxels per rotation in porous media experiments, we found RDL to remain effective up to approximately 14 voxels per frame in the capillary dataset. We attribute this to the ideal-case nature of the linear, vertical particle trajectories in the capillary, so that motion blur stretches particle images along the flow direction. In the capillary tube geometry, the measurements show that CTracks can recover particles with velocities up to approximately 75 voxels per frame. In porous media geometries, the more complex tracer trajectories induce stronger blurring effects which lead to more stringent limitations on the detectable velocities, as shown next.

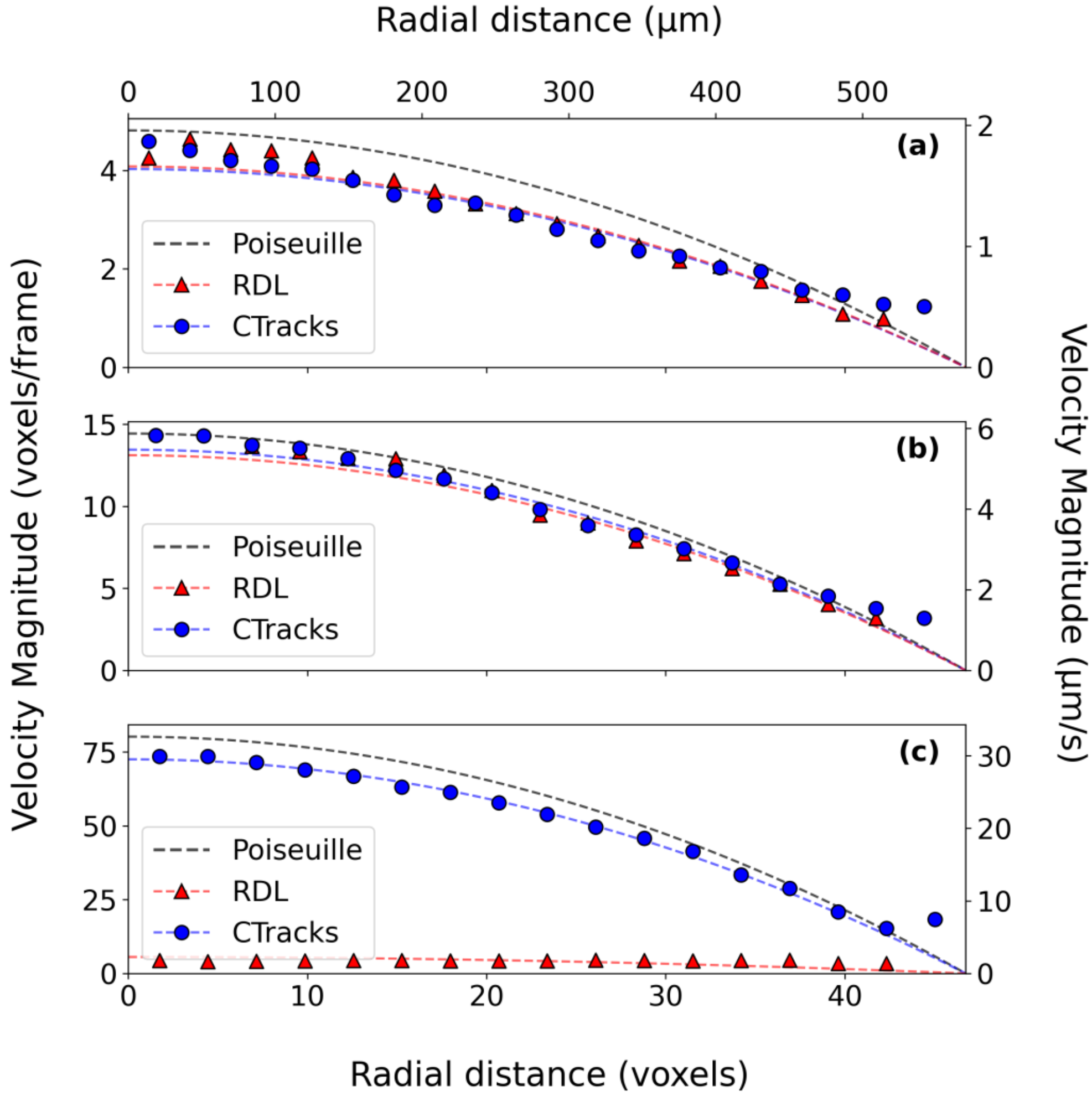

FIG. 4: Radial mean velocity magnitudes extracted from experimentally measured tracer trajectories in the capillary tube for each investigated flow rate. Results from RDL (conventional workflow) and CTracks (direct trajectory reconstruction) are plotted together with least-squares Hagen-Poiseuille fits. The theoretical velocity profile is plotted as a dashed black line. While both methods recover comparable profiles at 60 nl/min (a) and 180 nl/min (b), RDL breaks down at 1 µl/min (c) due to motion blur.

TABLE I: Relative errors in the measured interstitial velocity and fitted Poiseuille profile coefficient for the capillary-tube experiment. The profile coefficient was obtained by fitting the Hagen–Poiseuille equation to the measured radial velocity profiles. Although the fitted coefficients exhibit systematic bias at all flow rates, Fig. 4 shows that both methods recover similar profile shapes at 60 and 180 nl/min. At the highest particle displacement, RDL (conventional workflow) breaks down, whereas CTracks (direct trajectory reconstruction) maintains reasonable agreement with the expected velocity distributions.

| **Flow rate (nl/min)** | **Interstitial velocity error (%)** | | **Poiseuille profile coefficient error (%)** | |
|---|---|---|---|---|
| | RDL | CTracks | RDL | CTracks |
| **60** | -14 | -4.6 | -15 | -17 |
| **180** | -36 | 3.7 | -8.7 | -6.8 |
| **1000** | -98 | 16 | -93 | -9.5 |

### 2. *Porous glass*

In the porous-glass sample, the relative performance of the two methods depends strongly on flow velocity. As expected, the conventional RDL workflow performs well at the lowest flow rate of 60 nl/min. CTracks achieves comparable results, despite reconstructing fewer slow-moving particles. This behaviour is likely related to the current background-subtraction and particle-initialisation strategies, which reduce the detectability of slowly moving particles. While both steps are required components of the workflow, their present implementation is not fundamental to the direct trajectory reconstruction approach and could be improved in future versions.

At 60 nl/min, both methods provide a reasonable approximation of the estimated interstitial velocity reported in Table II. Some overestimation is expected as the interstitial velocity does not account for pore-space tortuosity and true mean pore velocities may therefore be one to two times higher than the nominal interstitial velocity (Fu et al., 2021). The comparable velocity estimates indicate that direct trajectory reconstruction remains effective in the low-velocity regime, before clear differences emerge at higher flow rates.

TABLE II: Relative error between the estimated interstitial velocity and the mean velocity in the flow direction, measured by each tracking algorithm in the porous-glass sample. CTracks (direct trajectory reconstruction) remains closer to the expected velocity as flow rate increases.

| **Flow rate (nl/min)** | **Interstitial velocity (vox/frame)** | **RDL relative error (%)** | **CTracks relative error (%)** |
|---|---|---|---|
| **60** | 0.7 | +14 | +23 |
| **180** | 2.1 | -73 | +3.3 |
| **1000** | 11.7 | -99 | -67 |

At the intermediate flow rate of 180 nl/min, corresponding to an approximately three-fold increase in interstitial velocity compared with previous XPTV experiments, clear differences emerge between the two approaches. While RDL reconstructs a larger population of slow or stationary particles, consistent with the discussion above, CTracks recovers substantially more fast-moving particles distributed throughout the pore space (Fig. 5). Many of these fast-moving particles are located along the preferential flow paths through the sample, which are only sparsely sampled by the conventional workflow.

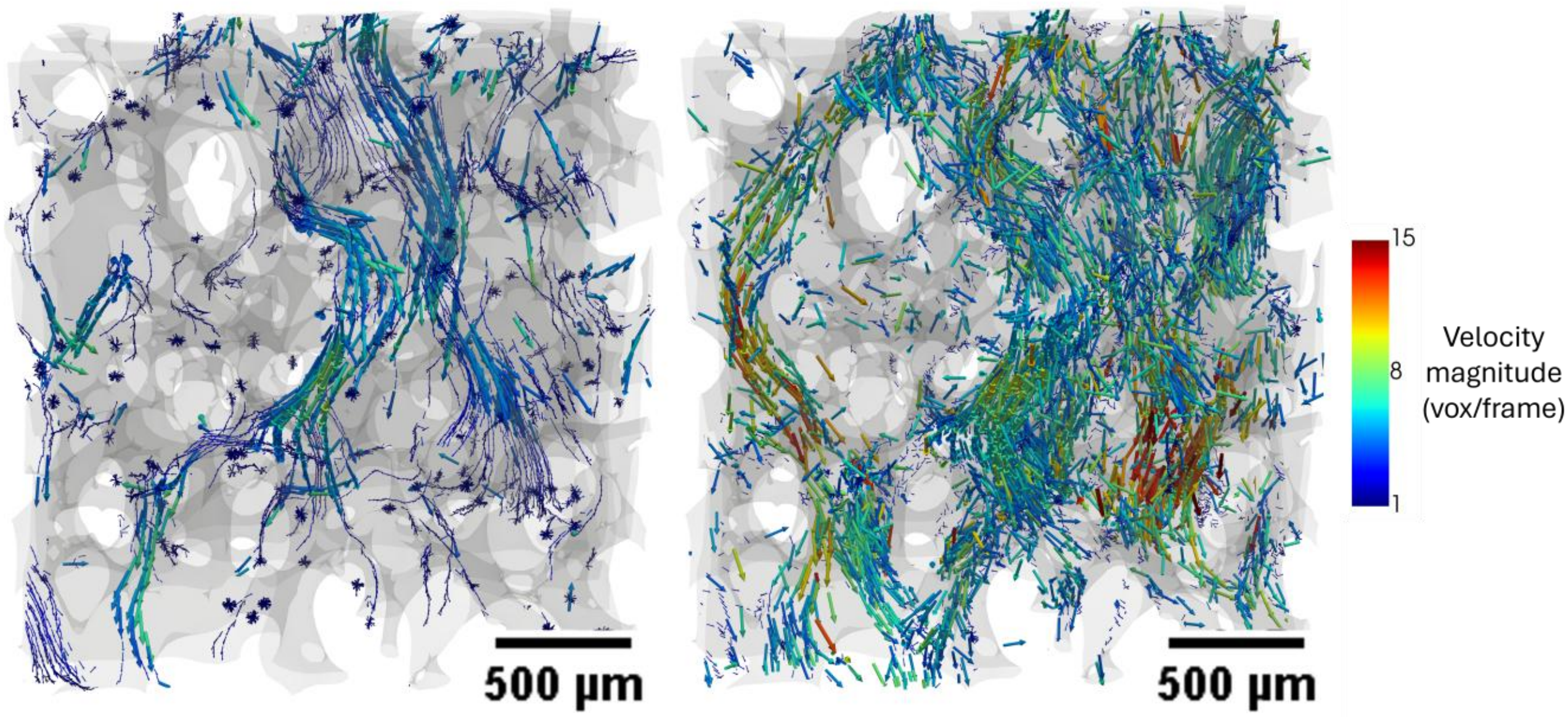

FIG. 5: Tracer velocity vectors measured in a zoomed 3D section of the porous-glass sample at 180 nl/min. Vectors are coloured by velocity magnitude. RDL (conventional workflow) is dominated by slow or stationary particles, whereas CTracks (direct trajectory reconstruction) recovers a larger population of fast-moving particles and improved sampling of the dominant flow pathways.

The velocity distributions shown in Fig. 6 support this interpretation. CTracks recovers many more high-velocity particles than RDL and produces a mean flow-direction velocity that closely matches the expected interstitial velocity (Table II). The velocity-magnitude distribution also shows a longer high-velocity tail, extending to approximately twice the velocity recovered by RDL. In contrast, RDL under-samples the faster particles and therefore underestimates the mean flow velocity. This improved agreement suggests that the additional high-velocity reconstructed trajectories correspond to genuine particle motion rather than measurement noise.

Figure 7 presents a qualitative comparison between the experimentally reconstructed flow fields and the corresponding CFD simulation described in Section III. Relative to RDL, the flow field reconstructed by CTracks more closely resembles the dominant high-velocity pathways predicted by the simulation. While the simulation is not intended as a ground-truth validation dataset, it provides a useful reference for the expected flow structure in this simple single-phase flow case. Consistent with this observation, the velocity-field correlation increases from $R = 0.25$ for RDL to $R = 0.61$ for CTracks at 180 nl/min (see Supplementary Information). Together with the particle-velocity distributions of Fig. 6, this suggests that the additional trajectories recovered by CTracks correspond to physically meaningful flow structures rather than reconstruction artefacts.

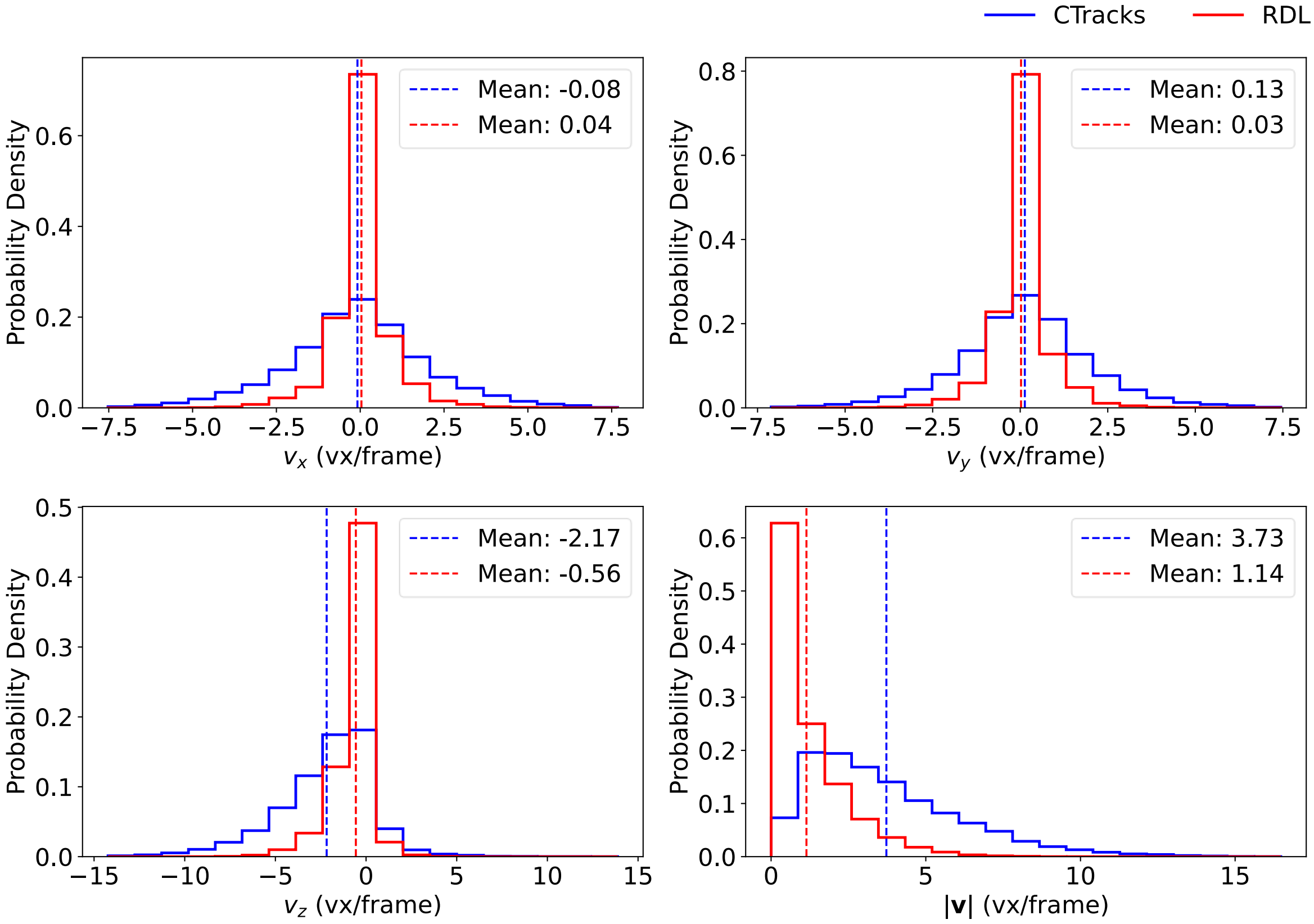

FIG. 6: Velocity distributions calculated from the experimentally measured tracer trajectories for the porous-glass sample at 180 nl/min. Mean values are indicated as dotted vertical lines. RDL (conventional workflow) under-samples the highest flow velocities, whereas CTracks (direct trajectory reconstruction) recovers a velocity distribution that more closely matches the expected interstitial velocity.

At the highest flow rate of 1 µl/min, RDL detects almost exclusively slow or stationary particles and provides little information about the dominant flow pathways within the sample (Fig. 8). At this flow rate, CTracks still recovers portions of the high-velocity flow structure, although many fast trajectories go undetected. Consistent with these observations, the correlation with the simulated flow field increases from R = 0.02 for RDL to 0.30 for CTracks (see Supplementary Information). Neither algorithm fully recovers the expected interstitial velocity at this flow rate, and this experiment can hence be seen as an upper bound of CTracks' current capabilities. Nevertheless, it also makes clear that CTracks outperforms RDL when motion blur limits the detectability of particles in the reconstructed CT volumes.

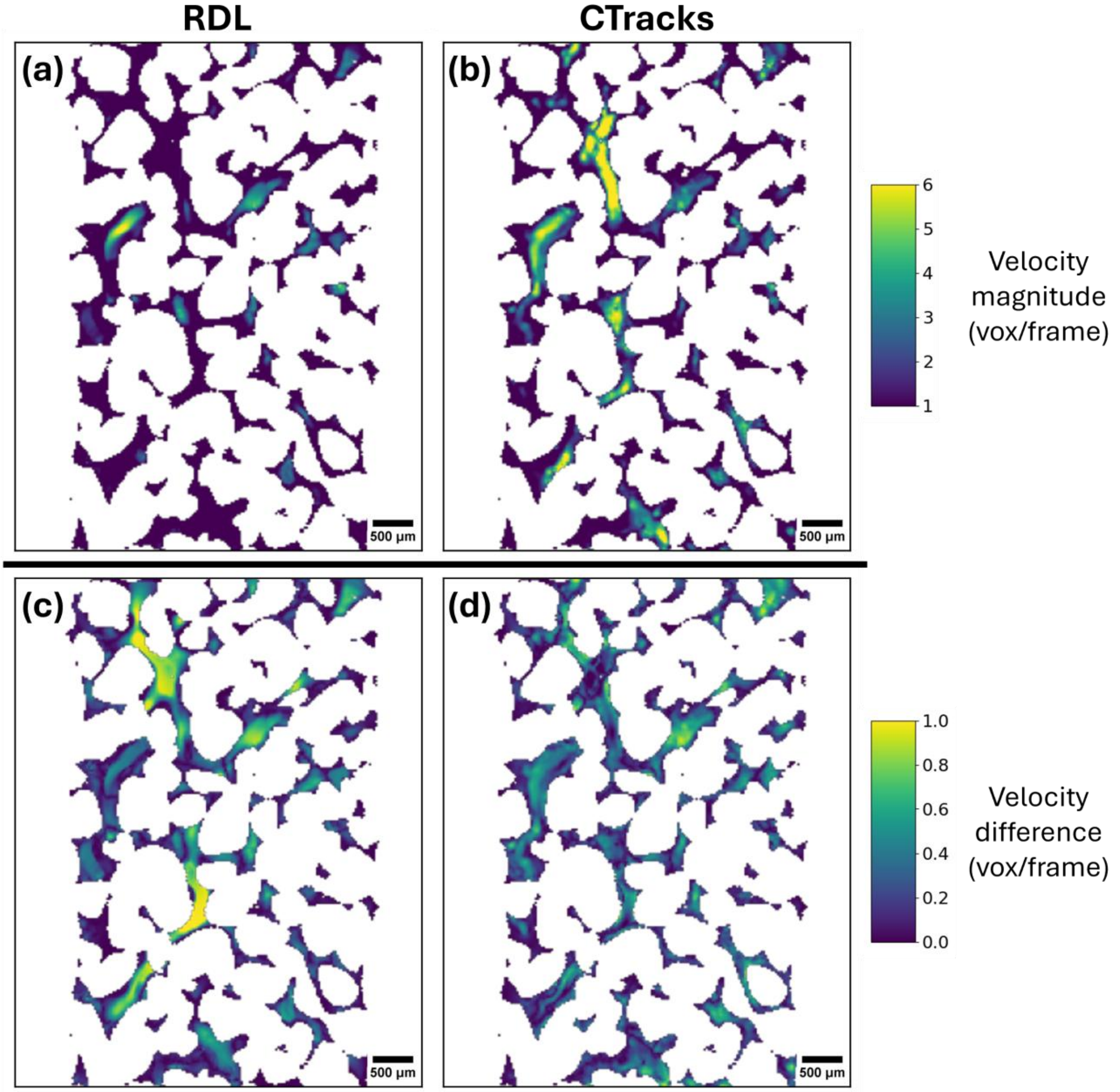

FIG. 7: Interpolated flow fields for the 180 nl/min experiment in the central vertical slice of the porous-glass sample. Measured velocity magnitudes are shown for (a) RDL (conventional workflow) and (b) CTracks (direct trajectory reconstruction). Corresponding absolute differences relative to the CFD simulation are shown for (c) RDL and (d) CTracks. CTracks shows improved agreement with the simulated flow field, particularly within high-velocity regions where particle motion is under-sampled by RDL.

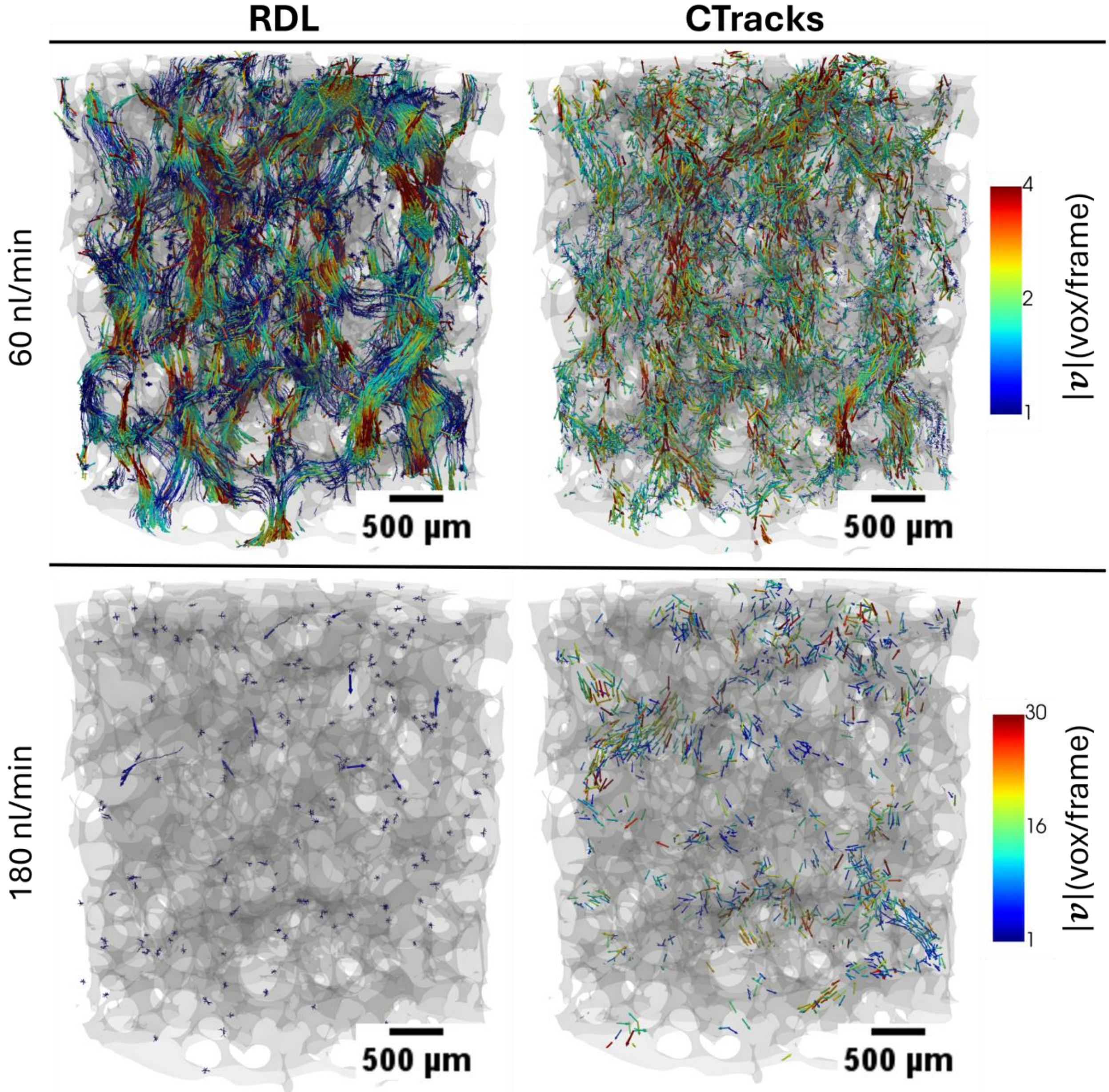

FIG. 8: Experimentally measured tracer trajectories reconstructed using RDL (conventional workflow) and CTracks (direct trajectory reconstruction) for the 60 nl/min (top) and 1 µl/min (bottom) flow rates in the porous-glass sample. At 60 nl/min, CTracks recovers fewer slow-moving particles than RDL, an effect attributed to the current background-subtraction and particle-initialisation strategy. At 1 µl/min, RDL detects primarily slow or stationary particles, whereas CTracks continues to recover portions of the high-velocity flow structure. Vectors are coloured by velocity magnitude (vox/frame).

### B. Validation simulation results

While the experimental datasets demonstrate improved performance of CTracks under realistic imaging conditions, the reconstruction accuracy cannot be quantified directly in experiments as the true particle trajectories remain unknown. We therefore use the simulation dataset introduced in Section III B to benchmark both tracking algorithms against known ground-truth trajectories. The simulated flow rate was deliberately selected above the experimentally investigated regime, allowing the performance limits of both methods to be evaluated. Because the synthetic projections were generated on an empty background and reconstructed using the same forward model, the simulated results are expected to overestimate absolute performance. They are therefore interpreted primarily in terms of relative trends between algorithms and with increasing velocity, rather than as a direct prediction of experimental performance.

Figure 9 shows recall, defined as the percentage of ground-truth particles successfully reconstructed, as a function of particle velocity magnitude for both tracking algorithms. RDL performs best for nearly stationary particles but rapidly loses detectability with increasing velocity and approaches zero at approximately 9 voxels/frame. In contrast, CTracks performs poorly for the slowest particles, a trend that was insensitive to both the trajectory-matching criterion and the removal of particles near grain boundaries, while remaining effective over a much wider velocity range. This behaviour is consistent with the observations of the experimental porous-glass dataset. Using a 20% recall threshold as a common reference point, the corresponding velocity increases from 5.5 voxels/frame for RDL to 27.6 voxels/frame for CTracks, representing an approximately five-fold increase of the detectable velocity range.

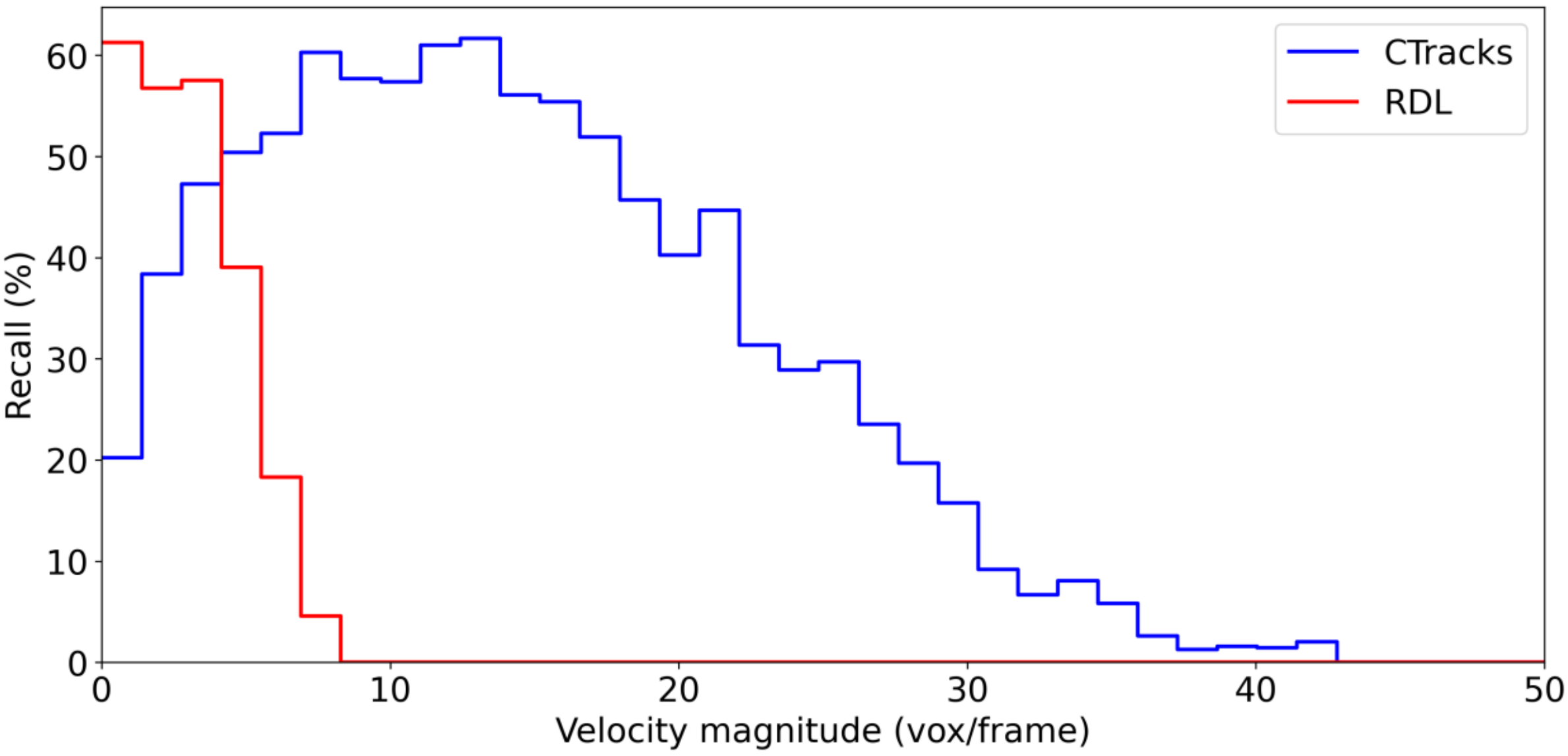

FIG. 9: Recall as a function of particle velocity magnitude for RDL (conventional workflow) and CTracks (direct trajectory reconstruction). RDL achieves slightly higher recall for the slowest particles, but recall decreases rapidly with increasing velocity. Using a 20% recall threshold as a common reference point, the corresponding velocity increases from 5.5 voxels/frame for RDL to 27.6 voxels/frame for CTracks, representing an approximately five-fold increase in detectable velocity range.

Extending the detectable velocity range is only useful if the recovered trajectories remain accurate. Figure 10 therefore evaluates the velocity magnitude error associated with successfully reconstructed CTracks particles. The velocity error remains approximately constant below 6 voxels/frame before gradually increasing, while remaining bounded by a relative error of less than 5% across most of the detectable range. The corresponding angular error exhibits a similar trend and is presented in Supplementary Fig. S2. Overall, the median error on CTracks velocities was 0.09 voxels/frame, while 90% of tracks had an absolute error below 0.69 voxels/frame.

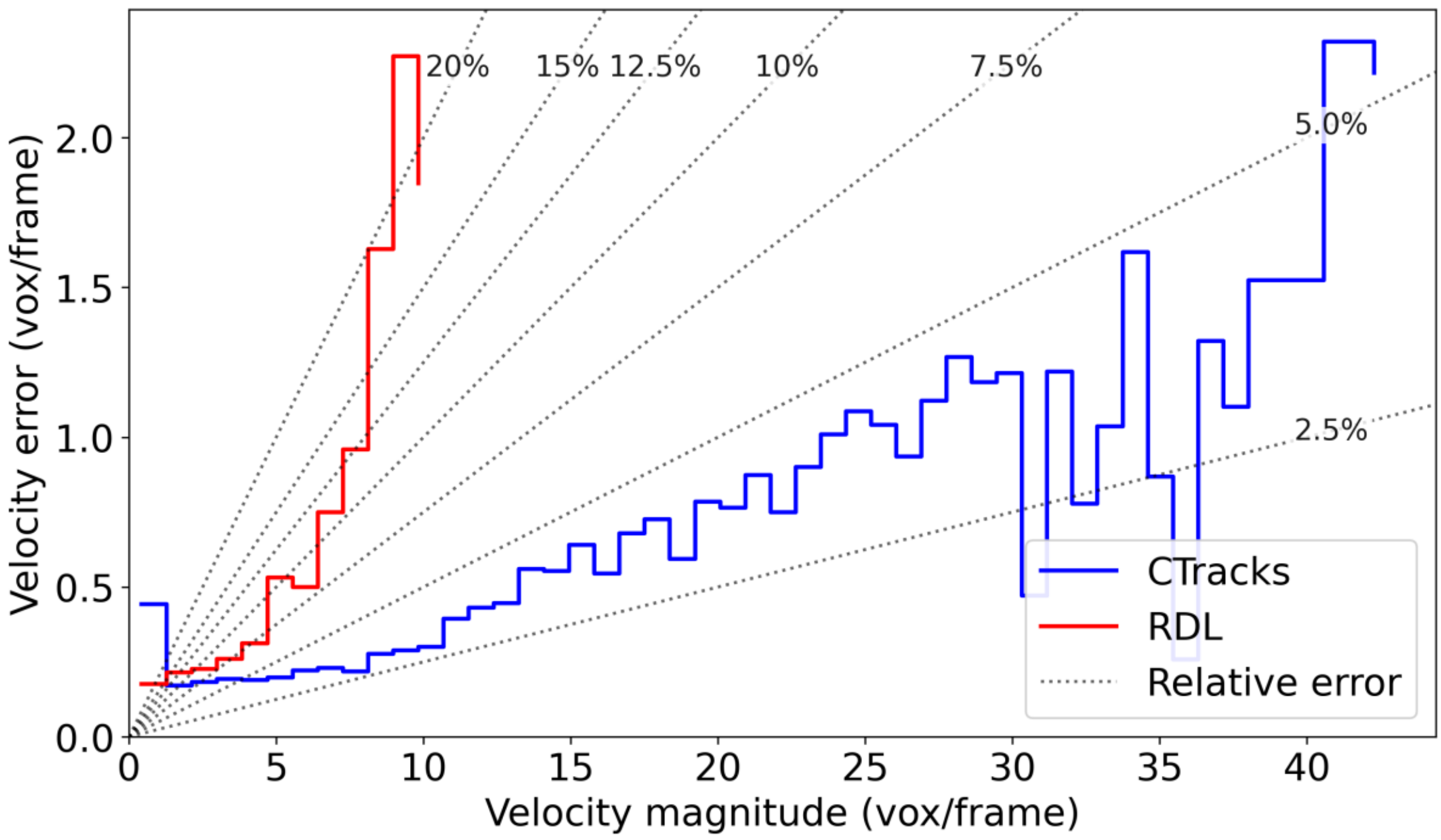

FIG. 10: Velocity magnitude error as a function of particle velocity magnitude for successfully reconstructed CTracks trajectories. The velocity error remains approximately constant below 6 voxels/frame before gradually increasing, while remaining bounded to less than 5% relative error across most of the detectable range.

In contrast, RDL remains accurate only over a much narrower velocity range. Within this range, RDL has slightly lower velocity errors for the slowest particles, consistent with its higher recall at low velocities. Similar trends are observed for the angular error (Supplementary Fig. S2). However, both errors increase more rapidly with particle velocity, reaching approximately 20% relative velocity error near the upper limit of its detectable range. Combined with the recall results of Fig. 9, this demonstrates that CTracks extends not only the velocity range over which particles can be detected, but also the range over which their velocities can be measured accurately.

Despite this substantial improvement in detectability, Fig. 9 shows that recall remains well below 100% across all velocities. To investigate the primary causes of this behaviour, we calculated detection persistence statistics for particles with velocities below 10 voxels/frame. As shown in Fig. 11, trajectories in this regime remain approximately linear, limiting the influence of the motion model on detection performance. A particle detected in one frame had a 94% chance of being detected in the following frame, whereas a missed particle had a 96% chance of remaining undetected. Since all simulated particles had the same attenuation and radius values, candidate particle initialisation appears to be a major factor limiting overall recall in the current implementation. Together with the background-subtraction strategy discussed previously, this likely explains the reduced recovery of slow-moving particles observed in both the experimental and simulated datasets. Detected particles are propagated between reconstruction intervals, implying increased likelihood of continued detection, whereas particles that are initially missed are rarely recovered later.

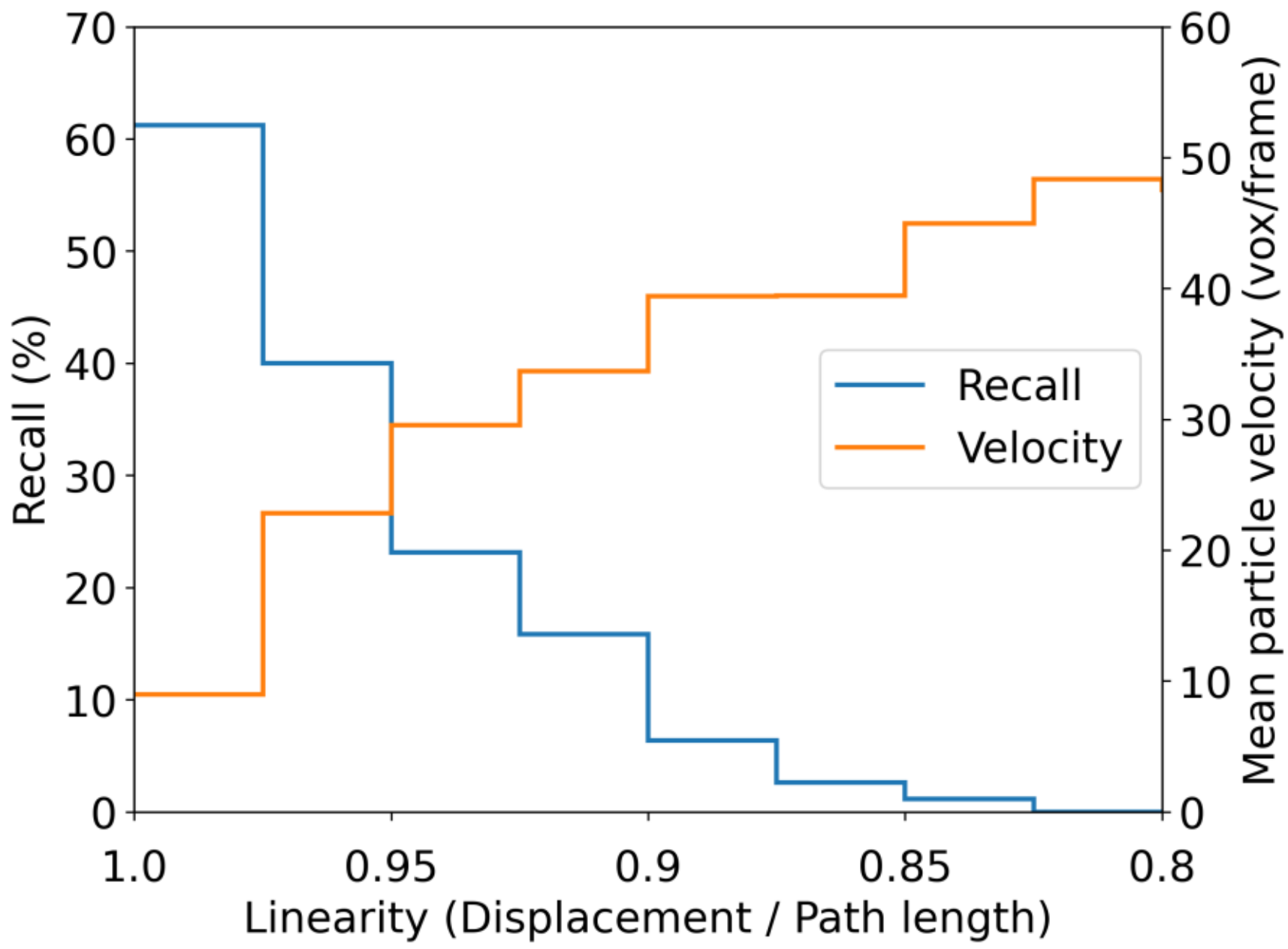

FIG. 11: Recall and mean particle velocity plotted as a function of trajectory linearity, defined as the ratio of the end-to-end displacement to the total path length over one CT rotation. A value of one corresponds to perfectly linear motion, while lower values indicate increasingly tortuous trajectories. Recall decreases with increasing tortuosity, which is also associated with higher particle velocities.

Figure 11 also relates recall to trajectory tortuosity, defined using the ratio of end-to-end displacement to total path length over a single CT rotation. A value of one corresponds to perfectly linear motion, while lower values indicate increasingly tortuous trajectories. Recall decreases with increasing tortuosity, suggesting that deviations from the linear motion parameterisation are an important limiting factor. This interpretation is consistent with the experimental capillary results, where particle motion was predominantly linear and CTracks remained effective up to much higher velocities. This suggests that the reduction in recall at high velocities is influenced not only by larger particle displacements, but also by increasing deviations from the linear trajectory parameterisation. Higher-order trajectory parameterisations are therefore expected to further improve performance at higher velocities.

Together, these analyses suggest two avenues for improving the current implementation. Improved particle initialisation strategies are expected to increase overall recall, while higher-order trajectory parameterisations should improve recovery of fast particles following increasingly tortuous paths. Both factors arise from implementation choices rather than limitations of the direct trajectory reconstruction methodology itself, suggesting substantial scope for further improvement.

## V. CONCLUSIONS

Tracking fast-moving tracer particles remains a key limitation of X-ray particle tracking velocimetry in porous media, restricting measurements to relatively slow flow regimes and preventing access to many realistic and highly dynamic pore-scale flow conditions. To overcome this limitation, we introduced a direct trajectory reconstruction approach in which particle trajectories are recovered directly from sequential radiographs rather than reconstructed CT volumes. By explicitly incorporating particle motion into the reconstruction process, this approach bypasses the motion-blur limitations that constrain conventional reconstruct-detect-link workflows. The concept was implemented through the CTracks framework and validated using both experimental and simulated datasets.

We first validated this method using experimental data from a capillary tube with known Poiseuille velocity profiles. We demonstrated excellent agreement between our new motion-aware iterative approach and conventional reconstruction methods at lower velocities (<14 voxels/frame). However, while the previous approach completely broke down and detected only noise at higher speeds, CTracks successfully recovered the expected parabolic velocity profile at particle displacements up to approximately 70 voxels/frame, a regime in which the conventional RDL workflow failed.

Following this proof-of-concept test, we applied CTracks to a complex porous-glass dataset acquired under flow conditions much faster than those previously demonstrated using XPTV (Bultreys et al., 2022). We observed significantly improved recovery of fast-moving particles and preferential flow paths that were poorly sampled or completely missed by the conventional workflow. This improvement was confirmed by comparing the experimentally measured flow field and the field generated through CFD simulation.

To benchmark the algorithms and perform error analysis, we constructed a digital twin of the porous-glass experiment, including both the pore-scale flow field and the X-ray imaging chain. The simulated flow conditions were selected to produce particle velocities approximately ten times higher than those observed experimentally, deliberately probing the performance limits of both tracking approaches. Using a 20% recall threshold as a common comparison point, the corresponding velocity range increased from 5.5 voxels/frame for RDL to 27.6 voxels/frame for CTracks, representing an approximately five-fold extension of the measurable velocity range. Additionally, the error from CTracks grew more slowly in both magnitude and angle, with velocity errors remaining approximately constant below 6 voxels/frame and bounded to less than 5% relative error across most of the detectable range.

While the detection of fast particles significantly improved, the performance decreased for slow particles (<6 voxels/frame) in both experimental and simulated data. Analysis of the simulated dataset suggests that this behaviour is primarily related to particle initialisation and the current background subtraction strategy rather than the direct trajectory reconstruction concept itself. Improved particle initialisation procedures and alternative background-estimation strategies are therefore expected to improve recovery of slow-moving particles in future implementations.

Future work will focus on refining the current implementation through improved particle initialisation strategies and higher-order trajectory parameterisations capable of representing the increasingly non-linear trajectories encountered at higher velocities. The simulation results indicate that both developments are likely to improve detection efficiency and further extend the attainable velocity range, particularly in highly tortuous geometries. In addition, optimisation of acquisition parameters specifically for trajectory reconstruction, rather than conventional CT reconstruction, may enable measurement of faster flows. Because detectability is primarily governed by particle displacement during a reconstruction interval, the combination of CTracks with the shorter acquisition times available at synchrotron facilities offers a promising route towards measurements of substantially faster pore-scale flows.

By bridging the gap between micrometre-scale spatial fidelity and high temporal resolution, direct reconstruction of particle trajectories from tomographic radiographs opens experimental access to a new regime of fluid dynamics in opaque media. Importantly, this expanded capability can be achieved using existing commercial micro-CT systems and conventional tomographic acquisition protocols, enabling direct measurements of faster pore-scale dynamics associated with multiphase flow instabilities and Haines jumps (Berg et al., 2013; Bultreys et al., 2024), as well as viscoelastic flow instabilities (Browne & Datta, 2021; Datta et al., 2022) under realistic experimental conditions. Ultimately, by capturing these pore-scale dynamics across an expanded dynamic range, this approach provides the experimental data needed to validate and refine continuum-scale flow models.

## APPENDIX

To illustrate the reconstruction of particles flowing through the experimental porous glass sample, a video of the central slice of the sample is shown in Fig. 12 (multimedia available online). Reconstructed particles from both RDL (red) and CTracks (green) are overlaid on each frame, with CTracks clearly detecting many more fast-moving particles.

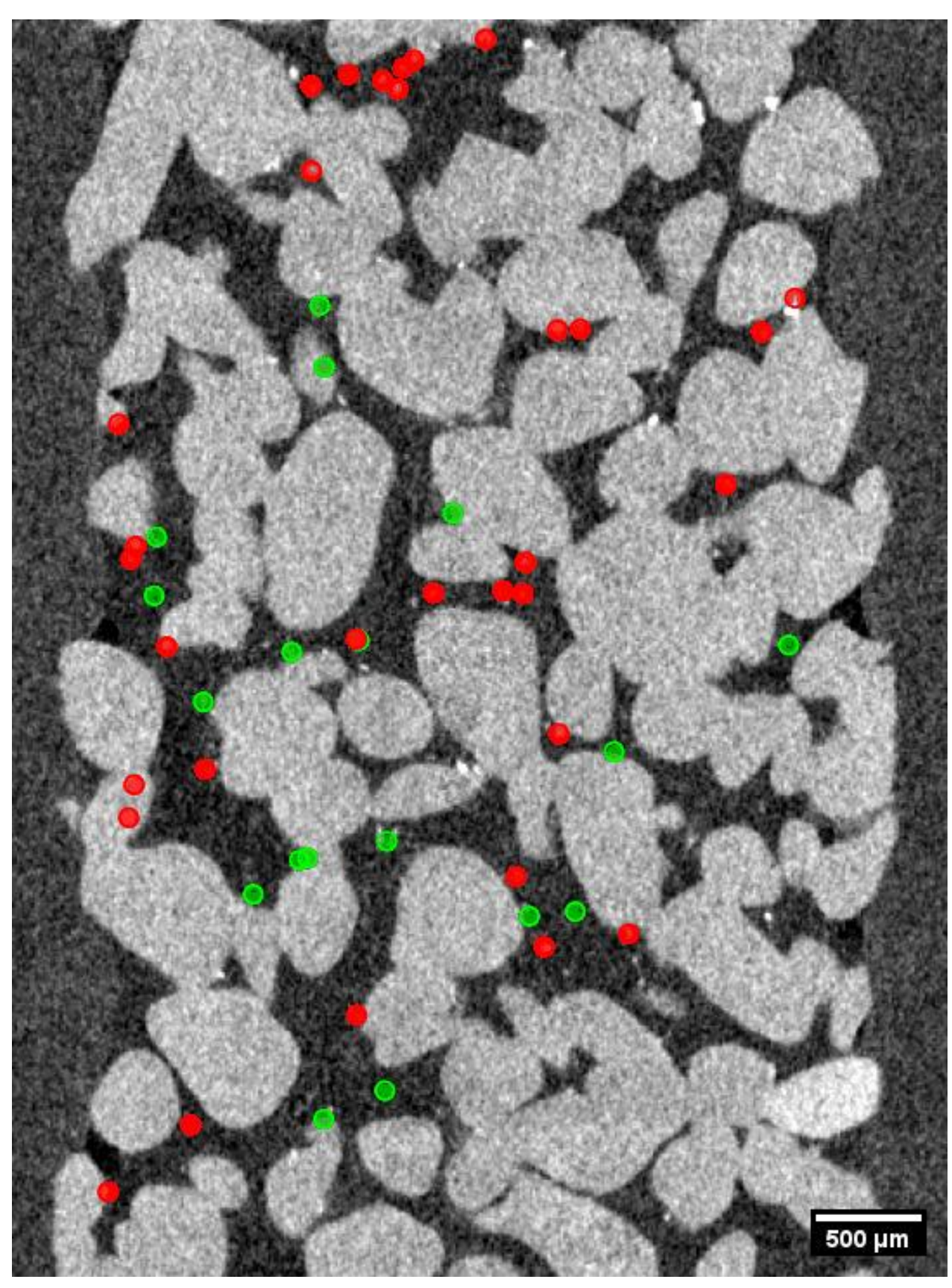

FIG. 12: Cross-sectional slices through reconstructed CT time frames of the porous-glass sample. Particle detections recovered by CTracks (green) and the conventional RDL workflow (red) are overlaid through time (Multimedia available online).

## ACKNOWLEDGEMENTS

Dr. Cyprien Soulaine is gratefully acknowledged for his guidance and mentoring in the use of OpenFOAM. The Ghent University Centre for X-ray Tomography (UGCT) is acknowledged for providing the experimental infrastructure used in this work. Support to the UGCT Core Facility from the Special Research Fund (BOF) of Ghent University (Grant No. BOF.COR.2022.008) is gratefully acknowledged. T.B. and R.v.d.M acknowledge support through the FWO-Tournesol programme (Grant No. VS00324N) which enabled a research visit to Dr. Cyprien Soulaine. This work was partially funded by the Research Foundation Flanders (FWO) under grant G0APF25N. This work is supported by ERC grant 101116228 (FLOWSCOPY). Funded by the European Union. Views and opinions expressed are however those of the author(s) only and do not necessarily reflect those of the European Union or the European Research Council Executive Agency. Neither the European Union nor the granting authority can be held responsible for them.

**DATA AVAILABILITY STATEMENT**

The CTracks reconstruction framework, trajectory analysis code and data that support the findings of this study are openly available in Zenodo. This includes the full experimental datasets containing the raw radiographic projections and the resulting particle trajectory data:

- CTracks framework: http://doi.org/10.5281/zenodo.22712231
- Capillary experiment: http://doi.org/10.5281/zenodo.22675888
- Porous glass experiment: http://doi.org/10.5281/zenodo.22662746
- Validation simulation: http://doi.org/10.5281/zenodo.22675822

**Direct Trajectory Reconstruction for Fast 3D X-ray Particle Tracking Velocimetry in Porous Media**

R. van der Merwe[1,2,a)], W. Goethals[1,2,3], S. Ellman[1,2], S. Manoorkar[1,2], J. Aelterman[1,4], M. N. Boone[1,3], and T. Bultreys[1,2]

[1)]Centre for X-ray Tomography, Ghent University, Proeftuinstraat 86, 9000 Ghent, Belgium
[2)]Department of Geology, Ghent University, Krijgslaan 281, 9000 Ghent, Belgium
[3)]Department of Physics and Astronomy, Ghent University, Proeftuinstraat 86, 9000 Ghent, Belgium
[4)]Department of Telecommunications and Information Processing – imec, Ghent University, Sint-Pietersnieuwstraat 41, 9000 Ghent, Belgium

## SUPPLEMENTARY INFORMATION

### S1. Experimental acquisition details

The experimental XPTV datasets introduced in Section III.A were acquired using the sample preparation and imaging procedures described below. The experimental setup is shown in Fig. S1.

The capillary tube was wrapped in polytetrafluoroethylene (PTFE) tape to improve mechanical stability and prevent leakage. Both the capillary and the sintered-glass sample were enclosed in Viton sleeves and mounted in an X-ray transparent Hassler-type flow cell (RS Systems, Norway). A confining pressure of 2 MPa was applied to prevent bypass flow.

The working fluid consisted of 85 wt% glycerol and 15 wt% deionised water, with partially hydrolysed polyacrylamide (HPAM) added to support parallel sample development and workflow studies. Although a polymer additive was included in the fluid formulation, under all imposed flow conditions the flow remained strongly viscous-dominated, with the Weissenberg number remaining below 1. The polymer was therefore not expected to influence the comparison between tracking algorithms.

Silver-coated hollow glass microspheres (Cospheric, USA) were used as tracer particles. These particles had nominal diameters of 5-22 μm and a density of 1.4 g/cm$^3$. These particles were dispersed in the glycerol-water base fluid at a concentration of approximately 8 mg/g. This mixture was stirred until visually dispersed, then sonicated and degassed before being combined with the HPAM solution. The polymer solution was not sonicated to avoid degradation of the polymer chains.

Before dynamic measurements, each sample was saturated with the unseeded working fluid using a syringe pump (Harvard PHD Ultra, USA). Approximately 4 ml of fluid was passed through each sample to displace trapped air. Two static scans were then acquired for each sample using the TESCAN CoreTOM at the Ghent University Centre for X-ray Tomography

[a)]Corresponding author, Robert.vanderMerwe@UGent.be

(UGCT). High-quality reference scans were acquired with a voxel size of 6.1 µm, 1400 projections, an integrated exposure time of 1.1 s per projection, a 1 mm Al filter, an accelerating voltage of 100 kV, and an X-ray source power of 15 W. Additional lower-quality static scans were acquired using the same imaging conditions as the dynamic XPTV measurements. These scans were used for the background subtraction step in the conventional RDL workflow, while the high-quality scans were used for pore-space segmentation. To obtain the pore-space segmentation used by both tracking workflows, grey-value thresholding was performed in Fiji, after which small, isolated features were removed using morphological opening, closing, and despeckle operations. The resulting binary volume was used to define the physically accessible regions of the sample for candidate particles.

All dynamic scans were acquired using scan settings of practical time-resolved XPTV, using a voxel size of 12.2 µm, 850 projections per rotation, an integrated exposure time of 35.25 ms per projection, an accelerating voltage of 60 kV, and an X-ray source power of 16 W. Under these conditions, one full tomographic rotation required 30 s. Tracer seeded fluid was injected using a Hamilton GasTight glass syringe mounted on a remote syringe pump (Harvard Nanomite, USA). Tracer arrival in the field of view was first confirmed by continuous two-dimension radiography, after which the flow rate was adjusted to the desired measurement condition and continuous tomographic acquisition was initiated. Dynamic datasets were acquired at volumetric flow rates of 60 nl/min, 180 nl/min and 1 µl/min. For the capillary, 40 rotations were recorded at 60 nl/min and 180 nl/min, and 20 rotations at 1 µl/min. For the sintered glass sample, 90 rotations were recorded at 60 nl/min, 60 rotations at 180 nl/min and 20 rotations at 1 µl/min.

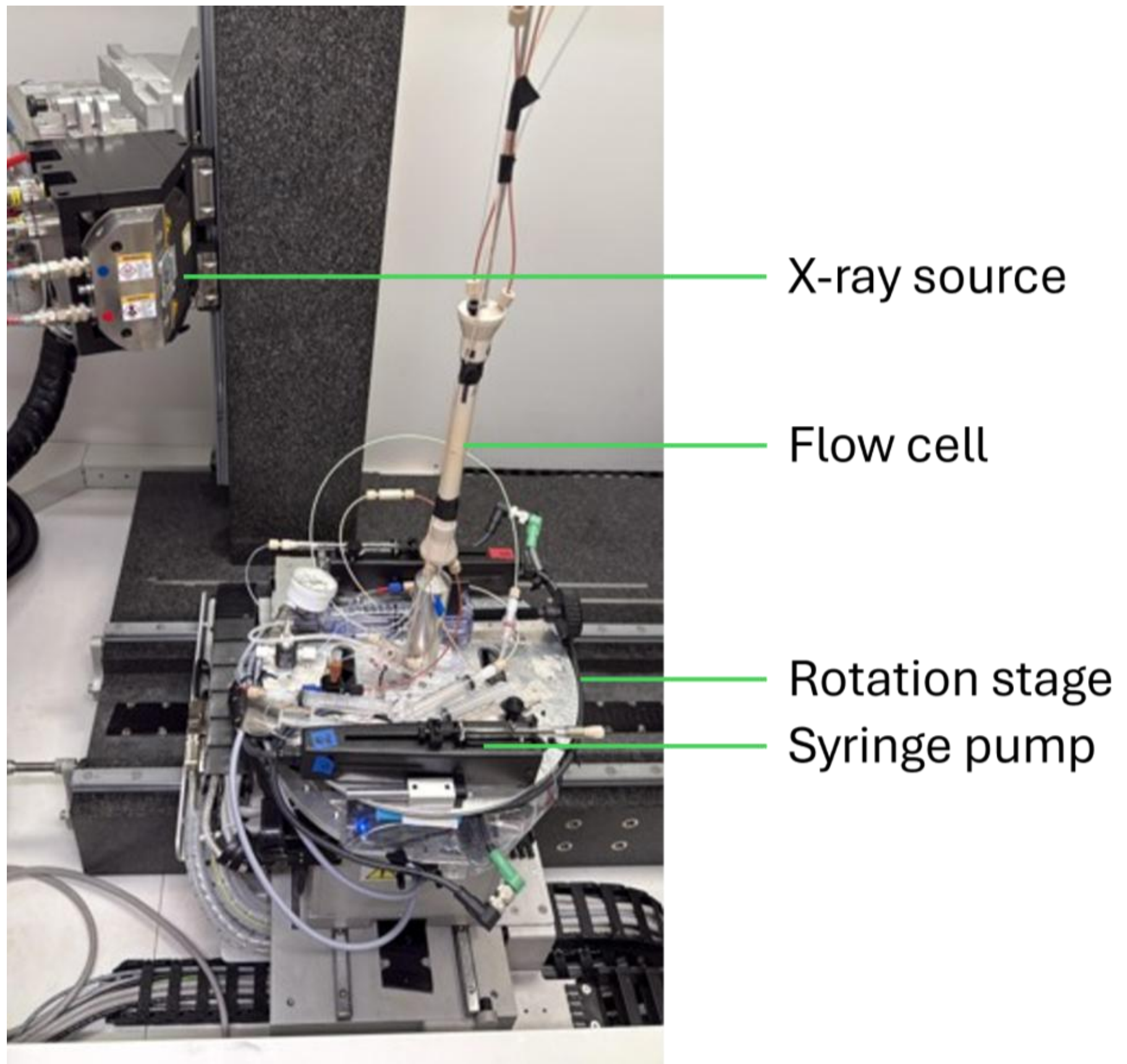

FIG. S1. Experimental XPTV setup mounted in the TESCAN CoreTOM system at the Ghent University Centre for X-ray Tomography (UGCT).

**S2. Projection-space implementation and preprocessing**

The mathematical formulation of the particle forward projector is given in Section II.B of the main manuscript. Here we summarise only the additional implementation details used in the present work.

Each tracer particle is represented as a small fixed-size patch on the detector (typically 7 × 7 pixels). This patch size is sufficient to capture the projected particle extent under the imaging conditions used, while maintaining computational efficiency. Because the particles are small relative to the source-detector distance, patch projection is implemented using a local small-angle approximation, with the patch dimensions scaled by the cone-beam magnification factor at the particle position. After computing the detector-space particle centre from the CT geometry, each patch is interpolated onto the detector grid and summed with all other particle contributions to form the generated projection at each projection angle. In addition to the particle parameters, selected acquisition geometry parameters can also be included in the optimisation, allowing compensation for calibration errors or scan instabilities. In this work, this was restricted to the centre-of-rotation.

To enable direct comparison between generated and measured data, the experimental radiographs were processed into the same background-subtracted optical-depth representation. Raw projection intensities were therefore converted to optical depth using the Beer-Lambert transform after which the static background was estimated directly from the dynamic dataset. The background estimation approach described in Section II.A assumes that the acquisition is sufficiently stable over the full duration of the experiment, with negligible bulk sample movement and limited beam fluctuation. When these assumptions are not satisfied, the same procedure can be applied within a sliding temporal window. While this improves robustness to slow drift, it also reduces the number of projections contributing to the minimum estimate and can therefore increase noise or partially remove slowly moving particles.

### S3. Optimisation settings

The CTracks framework is not inherently tied to a specific optimiser, batching strategy, or particle-cloud update schedule. The settings reported here correspond to the validation datasets presented in this work and should be regarded as implementation choices rather than fundamental components of the algorithm. Moderate variations in these settings, such as doubling optimisation epochs or halving batch sizes, produced similar reconstruction behaviour and did not qualitatively alter the conclusions presented here.

Trajectory parameters were optimised using the Adam optimiser. Optimisation was performed using batches of 20 projections, with projection angles randomly shuffled before batching to maximise angular diversity within each optimisation step. An iteration was defined as a single optimisation step performed on one batch, while an epoch corresponded to one complete pass through all projections associated with a reconstruction interval.

Each reconstruction interval was optimised for 100 epochs before commencing optimisation of the next interval. A learning-rate scheduler reduced the optimiser step size when no improvement in the global loss was observed over 20 consecutive epochs, halving all learning rates after each plateau event. To improve optimisation robustness and compensate for the bias introduced by the minimum-background estimation procedure, Gaussian noise with mean and variance estimated from particle-free regions of the experimental projections was added to the generated sinograms during optimisation.

For all datasets, optimisation of a single reconstruction interval (one CT frame) required approximately 5–8 minutes on an NVIDIA RTX A5000 GPU.

### S4. Particle initialisation and dynamic updates

Candidate particle parameters were initialised from empirically chosen distributions designed to match the expected tracer signal observed experimentally. In all datasets, a particle cloud of 600 candidate particles was used for initialisation. For both samples, particle radii were constrained between 0.8 and 2.5 voxels on the detector and initialised from a Gaussian distribution with a mean of 1 voxel and standard deviation chosen such that the full width at tenth maximum (FWTM) matched the allowed radius range. Particle attenuation values were constrained between 0.01 and 0.1 $\mathrm{cm}^{-1}$ and were initialised at the midpoint of this interval. Maximum allowed particle velocities were selected separately for each dataset. For the capillary dataset, the maximum velocity components $(v_x, v_y, v_z)$ were set to $(5, 5, 10)$, $(5, 5, 24)$, and $(5, 5, 120)$ voxels per timestep for flow rates of 60 nl/min, 180 nl/min and 1 μl/min respectively. For the porous glass dataset, the corresponding limits were $(5, 5, 10)$, $(7.5, 7.5, 15)$, and $(25, 25, 50)$ voxels per timestep.

Only particle positions were modified during the stochastic perturbation step. Perturbations were applied every 20 epochs, using an initial perturbation width of 3 voxels in each spatial direction. The perturbation width was reduced by 5% after each step. Any particle located within the segmented solid was relocated to the nearest pore voxel. Any particle that moved outside the vertical field of view during optimisation was removed, since they no longer contributed measurable signal and could otherwise vary arbitrarily without affecting the loss.

Additional candidate particles were introduced periodically during optimisation to account for tracer signals not represented by the current particle set. Particle insertion was evaluated every 30 epochs using the residual intensity difference between the generated and experimental sinograms. The average intensity contribution of a single particle trajectory was estimated by dividing the total generated sinogram intensity by the current number of candidate particles. To reduce the influence of image noise, the residual was first thresholded using a median absolute deviation (MAD) estimate obtained from particle-free image regions, retaining only residual pixels exceeding a 4σ threshold. Whenever the resulting residual exceeded 0.8 times the estimated mean single-particle trajectory intensity, additional particles were introduced according to the calculated excess signal. New particle positions were seeded randomly in the pore space, while initial velocity vectors were assigned from the mean velocity of the five nearest neighbouring particles. This threshold was intentionally selected to favour over-initialisation, which was found to improve reconstruction completeness compared to sparse initialisation strategies. Candidate particles were not explicitly removed during optimisation, apart from particles leaving the field of view, and remaining false detections were handled during post-processing.

## S5. Dataset-specific post-processing parameters

Post-processing was applied to improve the precision of the reconstructed trajectory set by removing implausible particle tracks while retaining physically meaningful velocity measurements. The post-processing workflow consisted of three sequential stages: feature-based filtering, Universal Outlier Detection (UOD), and spatial-isolation filtering. The dataset-specific post-processing parameters are summarised in Table S1. These parameters were not imposed globally as each dataset differed in particle density, background-subtracted image quality and velocity range. The listed values were therefore selected to improve the signal-to-noise ratio by reducing false particle detections while maintaining physically plausible velocity vectors.

Feature-based filtering thresholds were used to reject unlikely particle detections based on reconstructed particle size, attenuation, and velocity relative to the permitted maximum value. Universal Outlier Detection (UOD) was applied to identify locally inconsistent vectors by comparing each detection with neighbouring vectors. In this context, the residual threshold defines the maximum permitted deviation from the local neighbourhood motion after normalisation by the neighbourhood scatter, where lower values correspond to stricter rejection. Finally, spatial-isolation filtering was used to remove particularly sparse detections, as these were more likely either to be spurious or to provide insufficient local support to be useful. This filter could be applied either within individual timesteps or across all timesteps, depending on whether repeated detections of the same particle were expected.

The selected values reflect the expected flow structure and background-subtracted image quality of each dataset. In the capillary, the flow was smooth and spatially coherent, so moderate but uniform thresholds were sufficient across all conditions. In the porous glass dataset, at 60 nl/min motion was expected to be relatively slow and locally consistent, allowing slightly stricter filtering. At 180 nl/min, the broader distribution of velocities required a larger UOD neighbourhood and a more relaxed residual threshold, so that genuine variations in particle motion were not falsely rejected. The isolation filter was also applied per frame in this case. At 1 µl/min, extreme velocities led to highly non-linear particle motion during single CT frames, resulting in poor detectability. To extract the highest possible signal from this data, many thresholds had to be relaxed, ensuring the detection of some particles in increasingly linear regions of the geometry.

TABLE S1: Dataset-specific post-processing parameters used for trajectory quality control, including feature-based filtering thresholds, universal outlier detection (UOD) and spatial-isolation criteria.

| **Sample** | **Capillary** | **Porous glass** | | |
|---|---|---|---|---|
| **Flow rate** | **All** | **60 nl/min** | **180 nl/min** | **1 ul/min** |
| **Detection quality thresholds** | | | | |
| Radius (percentile) | 20 | 20 | 20 | 0 |
| Attenuation (percentile) | 30 | 30 | 30 | 20 |
| Distance to maximum velocity (%) | 10 | 10 | 5 | 10 |
| **Universal Outlier Detection** | | | | |
| Neighbourhood size | 5 | 5 | 10 | 5 |
| Residual threshold | 2 | 1.5 | 3 | 3 |
| **Spatial isolation filtering** | | | | |
| Isolation range (voxels) | 15 | 10 | 30 | 15 |
| Minimum neighbour count | 3 | 5 | 1 | 10 |
| Isolation per frame | No | No | Yes | No |

### S6. Simulated dataset generation

The porous geometry used for the synthetic validation dataset was reconstructed from a reference scan of the same sintered glass sample used experimentally. After grey-value thresholding in Fiji, binary erosion and dilation operations were used to reduce segmentation artefacts. The resulting fluid-solid interface was extracted in Python using a marching-cubes procedure and imported into OpenFOAM, where the pore space was discretised with snappyHexMesh, based on a $100^3$ background Cartesian mesh.

Steady single-phase incompressible Newtonian flow was solved in this geometry using a fluid density of 1.1 g/cm$^3$ and a kinematic viscosity of $2 \times 10^{-4}$ m$^2$/s. No-slip velocity and zero-gradient pressure boundary conditions were imposed at the grain surfaces. Flow through the sample was driven by a constant kinematic pressure difference of $27 \times 10^{-3}$ $m^2/s^2$, with zero-gradient velocity conditions at the inlet and outlet.

Tracer trajectories were then obtained by advecting ideal massless point particles through the Eulerian velocity field. An initial cloud of 5000 particles was distributed randomly throughout the pore space, and 15 additional particles were injected at the inlet each second. Particle positions were advanced with a time step of 0.04 s, selected to satisfy the Courant-Friedrichs-Lewy condition, and stored every 3 s.

To construct the synthetic radiographs, the discrete particle trajectories were interpolated in time to each projection angle using cubic interpolation and then forward projected using the experimental acquisition geometry. For the tracking analysis, only 1000 trajectories were used, chosen to avoid additional ambiguity associated with varying seeding density. In addition, all particles were enforced to remain within the field of view for the full simulated acquisition. In projection space, particle diameters were fixed to approximately 3.4 pixels and particle attenuation was fixed to $0.03\ \mathrm{cm}^{-1}$, corresponding to experimentally representative particle properties. This simplification neglects experimentally observed variability in particle size and attenuation but enables the isolation of trajectory-reconstruction performance from additional uncertainty associated with particle heterogeneity. Gaussian noise was added to reproduce the approximate noise level of the experimental radiographs.

As noted in Section III.B, the same forward projection model was used both to generate the synthetic data and to solve the inverse tracking problem. The synthetic benchmark should therefore be interpreted as an algorithmic validation under controlled conditions, rather than as a complete representation of all experimental uncertainties.

**S7. Quantitative comparison with simulated flow fields**

To complement the qualitative flow-field comparisons presented in the main manuscript, quantitative metrics were calculated between the simulated and interpolated experimental velocity fields. The mean absolute error (MAE) and root mean square error (RMSE) quantify differences in velocity magnitude, with RMSE placing greater emphasis on larger errors. The Pearson correlation coefficient (R) measures whether high- and low-velocity regions occur in the same locations, while the coefficient of determination ($R^2$) quantifies how much of the flow-field variability is explained by the reconstructed field. Table S2 summarises these metrics for each flow rate and tracking algorithm. While the magnitude-based metrics show minor differences between methods, much larger improvements are observed in the correlation-based metrics at higher flow rates. These results indicate that CTracks more accurately reproduces the spatial flow structure, particularly in regions where motion blur increasingly limits conventional RDL tracking.

TABLE S2: Quantitative comparison of experimentally measured and simulated velocity fields. Error-based metrics (MAE and RMSE) quantify differences in velocity magnitude, while correlation-based metrics (R and $R^2$) quantify the extent to which the measured field reproduces the simulated flow patterns. Lower MAE and RMSE values and higher R and $R^2$ values indicate improved agreement.

| **Flow rate (nl/min)** | **MAE** | | **RMSE** | | **Pearson R** | | **$R^2$** | |
|---|---|---|---|---|---|---|---|---|
| | RDL | CTracks | RDL | CTracks | RDL | CTracks | RDL | CTracks |
| **60** | 0.173 | 0.154 | 0.246 | 0.222 | 0.657 | 0.613 | 0.165 | 0.318 |
| **180** | 0.201 | 0.212 | 0.297 | 0.280 | 0.250 | 0.607 | -0.219 | -0.085 |
| **1000** | 0.184 | 0.191 | 0.321 | 0.302 | 0.025 | 0.296 | -0.465 | -0.297 |

**S8. Angular error analysis**

To complement the velocimetry magnitude error analysis presented in Fig. 10, Fig. S2 shows the angular error of successfully reconstructed trajectories. As with the magnitude error, angular accuracy is highest at lower particle velocities, decreasing with increasing velocity. However, the angular error remains bounded throughout the detectable velocity range, indicating that the reconstructed directions remain reliable.

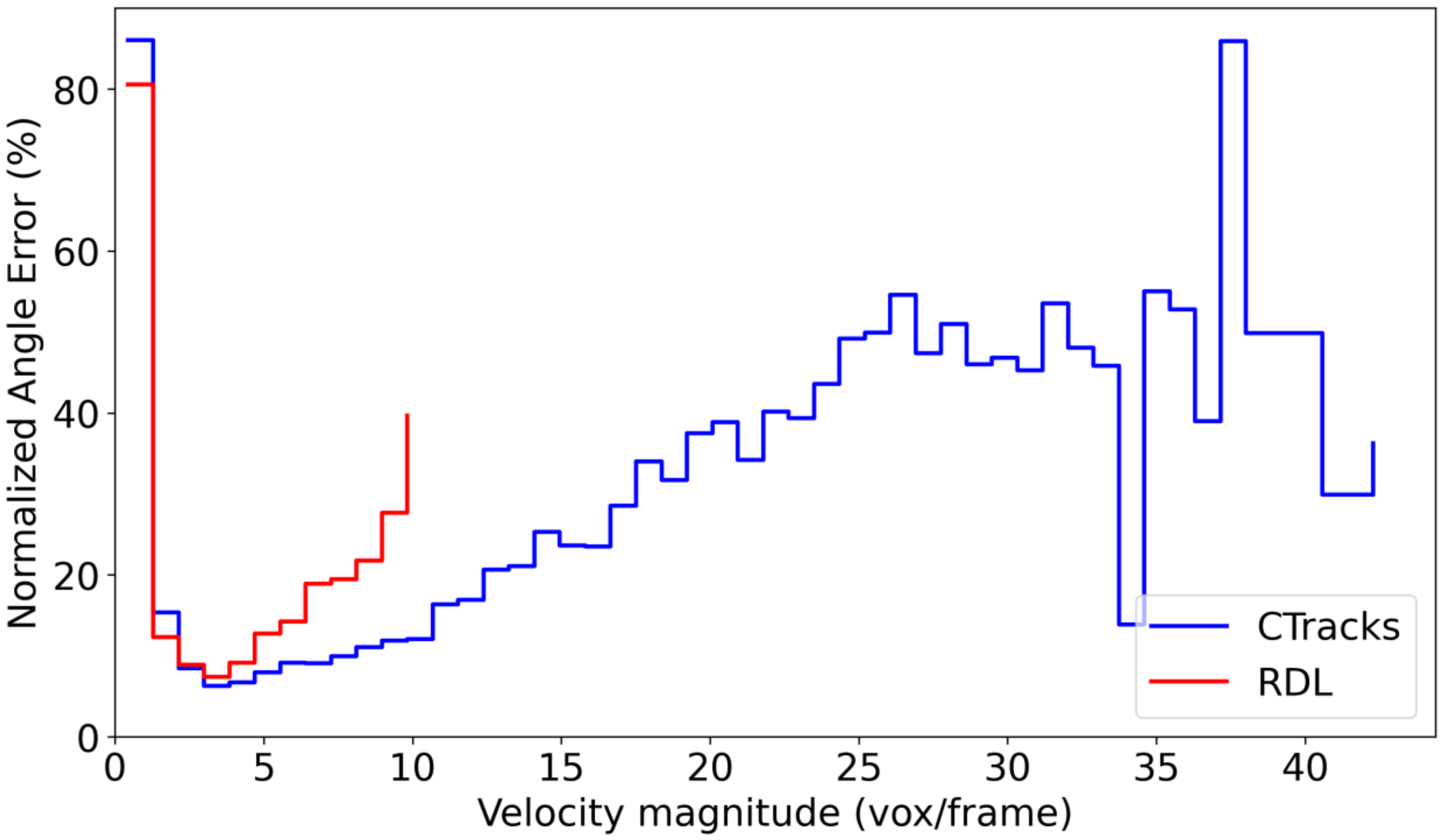

FIG. S2: Angular error as a function of particle velocity magnitude for successfully reconstructed trajectories. Angle errors are normalised by the maximum allowable angular deviation for a true positive, calculated as $\arcsin\left(\frac{2\sqrt{3}}{v}\right)$, where $2\sqrt{3}$ voxels is twice the voxel-diagonal endpoint tolerance used for trajectory matching. Angular error increases approximately linearly with velocity magnitude but remains bounded across the detectable velocity range.

## S9. Reconstruct-Detect-Link workflow

The reconstruct-detect-link (RDL) baseline followed the frame-based workflow of Bultreys et al. (2022), in which the measured radiographs were first reconstructed into a sequence of three-dimensional CT volumes, after which tracer particles were detected independently in each reconstructed frame and linked across time using the TrackPy toolkit. TrackPy is based on the Crocker-Grier particle-tracking framework, separating feature localisation from temporal linking. For particle detection, a pore-space mask derived from a registered dry reference scan was applied. Consistent with the workflow of Bultreys et al. (2022), the pore mask was subjected to two binary erosions and was used to remove false detections near pore walls caused by minor registration mismatches. A bandpass threshold of 1 was used during image pre-processing, after which candidate particles were identified using a particle-brightness percentile threshold of 98.5%, a feature size of 5 voxels, and a minimum feature separation of 4 voxels. Trajectories were then linked using a memory of 1 frame and a search range of 6 voxels. Only tracks with a minimum length of 20 frames were retained for subsequent analysis.